\documentclass[paper,notoc]{JHEP3}

\author{\footnote{KRYSTHAL collaboration} L.A. Harland-Lang$^1$, V.A. Khoze$^{2,3}$, M.G. Ryskin$^{2,4}$, W.J. Stirling$^{1,2}$\\ 
  $^1$Cavendish Laboratory, University of Cambridge,
  J.J.\ Thomson Avenue, Cambridge, CB3 0HE, UK\\
  $^2$ Department of Physics and Institute for Particle Physics Phenomenology, University of Durham, DH1 3LE, UK\\
$^3$ School of Physics \& Astronomy, University of Manchester,
Manchester M13 9PL, UK
$^4$ Petersburg Nuclear Physics Institute, Gatchina, St. Petersburg, 188300, Russia}

\title{Central exclusive meson pair production in the perturbative regime at hadron colliders}

\abstract{The central exclusive production (CEP) of heavy resonance states that subsequently decay into meson 
pairs, $M\overline{M}$, is an important signature for such processes at hadron colliders. However there is a potentially important 
background from the  {\it direct} QCD production of meson pairs, as mediated for example by the exclusive $gg\to M\overline M$
hard scattering subprocess. This is in fact an interesting process in its own right, testing novel aspects of
perturbative QCD technology. We explicitly calculate the $gg \to M\overline M$ helicity amplitudes for different meson states within the hard exclusive formalism, 
and comment on the application of MHV techniques to the calculation. Using these results, we describe how meson pair CEP can be calculated 
in the perturbative regime, and present some sample numerical predictions for a variety of final states. We also briefly consider the dominant non-perturbative contributions,
which are expected to be important when the meson transverse momentum is small.}

\preprint{IPPP/11/19\\  DCPT/11/38\\ Cavendish-HEP-11/06}

\usepackage{cite}
\usepackage{subfigure}
\usepackage{multirow}
\usepackage{helvet}
\usepackage{amsmath}
\usepackage{amssymb}
\usepackage{setspace}
\usepackage{setspace}
\usepackage[dvips]{graphicx}
\usepackage{epsfig}
\def\lesim{ \;\raisebox{-.7ex}{$\stackrel{\textstyle <}{\sim}$}\; }
\def\be{\begin{equation}}
\def\ee{\end{equation}}

\begin{document}

\section{Introduction}

Central exclusive production (CEP) processes of the type
\begin{equation}\label{exc}
pp({\bar p}) \to p+X+p({\bar p})\;,
\end{equation}
can significantly extend the physics programme at high energy hadron colliders. Here $X$ represents a system 
of invariant mass $M_X$, and the `$+$' signs denote the presence of large rapidity gaps. 
Such reactions provide a very promising way to investigate both QCD dynamics and new physics in hadron collisions, 
and consequently they have been widely discussed in the literature (see~\cite{KMRprosp,fp420,Royon:2008ff,afc,Albrow:2010zz} for reviews).
Of special interest is the study of resonance states, from `old' SM mesons to BSM Higgs bosons (see for example 
\cite{Kaidalov03,Khoze:2004rc,HarlandLang10,Antoni,HKRSTW,shuvaev} and references therein). 
An attractive advantage of CEP reactions is that they provide an especially clean environment in which to test the nature and the quantum numbers of the centrally produced state $X$.

Recall (see~\cite{KMRprosp,Khoze:2000mw,Khoze00a}) that in exclusive processes, as shown in Fig.~\ref{fig:pCp}, the incoming $gg$ state satisfies 
special selection rules in the limit of forward outgoing protons, namely it has $J_z = 0$, 
where $J_z$ is the projection of the total $gg$ angular momentum on the beam axis, and positive $C$ and $P$ parity.  
Hence only a subset of the helicity amplitudes for the $gg \to X$ sub-process contributes. The CEP mechanism therefore provides a 
unique possibility to test the polarization structure of the $gg \to X$ reaction.

Even without forward proton spectrometers, double diffractive processes of the type
\begin{equation}\label{dd}
pp({\bar p}) \to Y+X+Z\;,
\end{equation}
with large rapidity gaps separating the centrally produced system $X$ from the products, $Y$ and $Z$, of the proton 
(antiproton) dissociation, are still of considerable interest. If the incoming protons are dissociated into low masses ($M_{Y,Z}\lesim$ 2 GeV), 
then these reactions continue to exhibit many of the attractive properties of CEP, while in the case of high-mass dissociation the $J_z=0$ selection rule, characteristic of exclusive production, no longer holds\footnote{See footnote 3 after (\ref{simjz2}).}.

These double diffractive processes can be measured using forward rapidity gap triggers, with the help of simple scintillation 
(forward shower) counters  (FSCs) \cite{fsc}. Such a strategy was successfully implemented at the Tevatron by the CDF collaboration 
\cite{Albrow:2010zz
, 
Abulencia:2006nb,cdf:2007na,Aaltonen:2009kg,Aaltonen:2009cj}, where the rapidity gap trigger was used to veto on particles with pseudorapidity
$|\eta| <$ 7.5 on each side of the central system. 
CDF  have published a search for $\gamma \gamma$ CEP \cite{cdf:2007na} with $E_T(\gamma) >$ 5 GeV. 
This process (together with the CEP of charmonia, the observation of which was reported in~\cite{Aaltonen:2009kg}), can serve as a 
`standard candle' reaction with which we can check the predictions for new physics CEP at the LHC \cite{HarlandLang10,Khoze:2004ak}. 
Based on the results in~\cite{Khoze:2004ak}, the observed CDF $\gamma \gamma$ CEP cross section should be small, 
corresponding to 0.8$^{+1.6}_{-0.5}$ events; experimentally~\cite{cdf:2007na}, three candidate events were observed. 
Subsequently the $E_T(\gamma)$ threshold has been decreased to $\sim 2.5$ GeV,
 and  many more candidate events have been observed \cite{Albrow:2010zz,Albrow}. 
In this context, a good quantitative theoretical understanding of the  CEP $\pi^0\pi^0$ background is crucial, since one or both of the 
photons from $\pi^0 \to \gamma\gamma$ decay can mimic the `prompt' photons from $gg \to \gamma\gamma$ CEP.
 
As discussed in~\cite{HarlandLang10,Khoze04,HarlandLang09}, the observation of $\chi_{c0}$ CEP via two-body decay channels 
is of special interest for both studying the dynamics of heavy quarkonia and for testing the QCD framework of CEP, with 
the $\chi_c \to \pi\pi$ decay being a promising example. We recall that these channels, especially $\pi\pi$, $K^+K^-$ and $p\bar{p}$, 
are ideally suited for spin-parity analysis of the $\chi_c$ states. In particular, the fact that the $\chi_{c(1,2)}$ two-body 
branching ratios are in general of the same size or smaller (or even absent for the $\chi_{c1}$) than for the $\chi_{c0}$, ensures 
that the $J_z=0$ selection rule is fully active, see~\cite{Khoze:2000mw,Khoze04,HarlandLang09} for more details. However, in this case we may  
expect a sizeable background resulting from the direct QCD production process
\begin{equation}\label{hh}
pp({\bar p}) \to p+h\bar h+p({\bar p})\;
\end{equation}
with $h=\pi,K,p$. Such a non-resonant contribution should therefore be carefully evaluated. We recall that  the existing CDF 
measurement of $\chi_{c}$ CEP \cite{Aaltonen:2009kg} is based on the detection of the decay $\chi_c \rightarrow J/\psi+\gamma$.
While the overall observed rate is in  agreement with theoretical
expectations~\cite{Khoze04,HarlandLang09}, the M($J/\psi+\gamma$) resolution, due to the low photon energy ($\sim 200$~MeV), 
does not allow a separation of the different $\chi_{cJ}$ states. 
Therefore, although the $\chi_{c0}$ should be the dominantly (CEP) produced $\chi_c$ state, the higher spin  
$\chi_{c1}$ and $\chi_{c2}$ states could give a sizeable contribution to the observed  $J/\psi+\gamma$ signal, 
because of their much higher branching fractions to $J/\psi+ \gamma$~\cite{HarlandLang09,teryaev}.

While the Tevatron CEP data do not in general make use of forward proton detectors, relying 
instead on large rapidity gap triggers,
 a new area of experimental studies of CEP with {\it tagged} forward protons is now being explored by the STAR Collaboration 
at RHIC \cite{wlodek}, which
has the capability to trigger on and to measure forward protons. This provides an excellent means to extend the physics reach 
in studying CEP processes.

Currently at the LHC, without forward proton taggers, the existing detectors (ALICE, ATLAS, CMS and LHCb) were not designed 
for studying such standard candle processes as charmonia (e.g. $\chi_{c}$) or $\gamma \gamma$ CEP, because they 
lack  the forward coverage necessary to measure large rapidity gaps. Selecting only events with a low multiplicity 
in the {\it central} detector may not be enough to study these reactions since such events include processes with  
proton dissociation, as in (\ref{dd}) 
However, as mentioned above, the addition of FSCs could provide sufficient 
rapidity coverage and would allow the exclusion of events with high-mass and a large fraction of events 
with low-mass diffractive dissociation. Thus events with a `veto' FSC trigger (assuming that the number of interactions
per bunch crossing is not too large) may be interpreted as arising from CEP. In fact such counters have been proposed for CMS
\cite{fsc,cms} and are currently being installed. This will open up a wide physics programme at CMS based on vetoes
 using the FSC, ZDC, CASTOR and HF detectors, together with requiring some minimal activity in the 
central region\cite{Albrow:2010zz,fsc,cms,orava}. In particular, CEP events of the type (\ref{exc}) can be 
selected without detecting forward  protons.

Finally, we note that by employing FSCs a rich CEP physics programme could also be realised with the 
LHCb experiment~\cite{lhcb}. The excellent particle identification of the LHCb detector and the high 
momentum resolution are especially beneficial for measurements of low multiplicity final states\footnote{Recently the LHCb collaboration have reported  encouraging preliminary results on possible CEP in the $\chi_{c(0,1,2)}\to J/\psi\,+\,\gamma$ channel, where in order to justify the exclusivity vetoing was imposed on additional activity in the rapidity region $1.9<\eta<4.9$, with some sensitivity to charged particles in the backwards region $-4<\eta<-1.5$~\cite{lhcbexc}. While the $\chi_{c(0,1)}$ production data are in good agreement (within theoretical and experimental uncertainties) with our predictions for CEP~\cite{HarlandLang09,HarlandLang10}, the observed $\chi_{c2}$ rate is somewhat higher. However, as discussed above the observed data will include events with proton dissociation, as in (\ref{dd}). Such an inclusive contamination, where the momentum $p_\perp$ transferred through the $t$-channel  $gg$  system  (see footnote 3) may be rather large, will tend to preferentially increase the higher spin $\chi_{c(1,2)}$ yield. Thus the $\chi_{c2}$ cross section, which  is  proportional to $p_\perp^4$, could be particularly enhanced. This may therefore account for some or all of the observed disagreement between the LHCb $\chi_{c2}$ measurement and the CEP theory. Although a data-driven attempt to subtract the inclusive contribution has been made, it is not clear that this will account for all of the background: as discussed above, the addition of FSCs on both sides of the LHCb experiment would allow a more efficient veto on such inelastic events~\cite{lhcb}, and would therefore greatly clarify the situation.}.
A promising study of low central mass CEP events is also ongoing at ALICE \cite{rainer}, using additional scintillator detectors placed 
on both sides of the central barrel, which allows tagging of double rapidity gap events.

Studies of meson pair CEP would also present a new test of the perturbative formalism, with all 
its non-trivial ingredients, from the structure of the hard sub-processes to the incorporation of rescattering effects 
of the participating particles. Of particular interest is $\eta\eta$ and $\eta'\eta'$ CEP, which could allow a probe of 
the gluonic structure of $\eta, \eta'$ mesons, the available theoretical information on which is still quite scarce. 
Though being undoubtedly of interest in their own right, the two-meson non-resonant exclusive processes (\ref{hh}) are 
of essential practical importance as the main backgrounds to  $\chi_c \to \pi\pi, KK$ CEP, and, in the 
case of $\pi^0\pi^0$, to $\gamma\gamma$ CEP. This in turn requires a very detailed theoretical knowledge of the 
properties of the process (\ref{hh}).

$\pi\pi$ CEP, mediated by Pomeron-Pomeron fusion, has been the subject of theoretical studies within a Regge-pole framework 
since the mid-1970s (see, for example \cite{Azimov74,Desai78} for early references and 
\cite{Lebiedowicz09,HarlandLang:2010ys} for more recent ones). There have also been a variety of experimental results on 
low-mass meson pair CEP, in particular from the CERN ISR, with the measured cross sections in broad agreement with the 
expectations of Regge phenomenology (see~\cite{afc,Albrow:2010zz} for reviews). As discussed in \cite{HarlandLang:2010ys}, 
at comparatively large meson transverse momenta, $p_\perp$, CEP should be dominated by
the perturbative 2-gluon exchange mechanism of Fig.~\ref{fig:pCp}, where the $gg\to \pi \pi,K^+K^-$ coupling can be calculated using the
formalism of~\cite{Brodsky81,Benayoun89}. In the kinematic regime relevant for $\chi_c$ CEP 
($M_{\pi\pi}\sim M_\chi$, $p_\perp(\pi)\sim M_\chi/2$), both non-perturbative {\it and} perturbative mechanisms could in principle  
contribute to the overall rate, and this issue requires careful investigation. 
Of special interest is the transition region between these two contributions, which is very sensitive to the 
behaviour of the meson form factor $F_M(t)$. The potential $\pi\pi$ CEP continuum background to $\chi_{c0}$ was also recently considered purely within the framework of Regge theory by~\cite{Lebiedowicz11} (see also~\cite{Staszewski:2011bg}) -- we will briefly comment on and compare our results in Section~6.

In this paper we perform a detailed study of the CEP of meson pairs, paying special attention to the perturbative regime. 
First, in Section~2, we review the CEP calculation formalism for processes (\ref{exc}). We then in Section~3 consider meson pair, $M\overline{M}$, production, in particular the perturbative QCD calculation of the $gg \to M\overline{M}$ process, thus generalising the existing $\gamma\gamma \to M\overline{M}$ results. Unlike the $\gamma\gamma$ case,
the $gg \to M\overline{M}$ cross section exhibits an interesting (QCD) radiation zero. We also show how 
MHV techniques can be used to calculate the underlying $gg\to q\bar q q\bar q$ helicity amplitudes. 
In Section~4 we address the calculation of $\eta\eta$,  $\eta\eta'$ and $\eta'\eta'$ production, and in Section~5 we 
calculate and compare the subprocess and hadron-level cross sections for various final states. Non-perturbative meson pair CEP is 
considered in Section~6, and a secondary perturbative mechanism is described and compared with our standard mechanism 
in Section~7. Finally, Section~8 contains our conclusions and outlook for further work.

\section{Central exclusive production}\label{CEPform}

\begin{figure}[b]
\begin{center}
\includegraphics[scale=1.0]{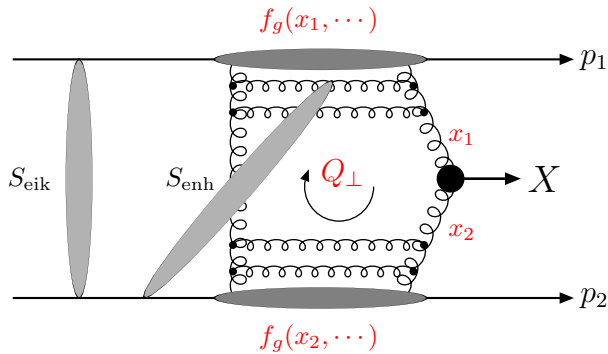}
\caption{The perturbative mechanism for the exclusive process $pp \to p\,+\, X \, +\, p$, with the eikonal and enhanced survival factors 
shown symbolically.}
\label{fig:pCp}
\end{center}
\end{figure} 

The formalism used to calculate the perturbative CEP cross section is explained in detail elsewhere~\cite{Kaidalov03,HarlandLang10,Khoze00a,Khoze00} and so we will only review the relevant aspects here. The amplitude is described by the diagram shown in Fig.~\ref{fig:pCp}, where the hard subprocess $gg \to X$ is initiated by gluon-gluon fusion and the second $t$-channel gluon is needed to screen the colour flow across the rapidity gap intervals. We can write the `bare' amplitude in the factorised form~\cite{KMRprosp,Khoze04,KKMRext,Khoze:2004ak}
\begin{equation}\label{bt}
T=\pi^2 \int \frac{d^2 {\bf Q}_\perp\, \mathcal{M}}{{\bf Q}_\perp^2 ({\bf Q}_\perp-{\bf p}_{1_\perp})^2({\bf Q}_\perp+{\bf p}_{2_\perp})^2}\,f_g(x_1,x_1', Q_1^2,\mu^2;t_1)f_g(x_2,x_2',Q_2^2,\mu^2;t_2) \; ,
\end{equation}
where the $f_g$'s in (\ref{bt}) are the skewed unintegrated gluon densities of the proton: in the kinematic region relevant to CEP, they are given in terms of the conventional (integrated) densities $g(x,Q_i^2)$. $t_i$ is the 4-momentum transfer to proton $i$ and $\mu$ is the hard scale of the process, taken typically to be of the order of the mass of the produced state: as in~\cite{Khoze04,KKMRext}, we use $\mu=M_X/2$ in what follows. The $t$-dependence of the $f_g$'s is isolated in a proton form factor, which we take to have the phenomenological form $F_N(t)=\exp(bt/2)$, with $b=4 \,{\rm GeV}^{-2}$. The $\mathcal{M}$ is the colour-averaged, normalised sub-amplitude for the $gg \to X$ process
\begin{equation}\label{Vnorm}
\mathcal{M}\equiv \frac{2}{M_X^2}\frac{1}{N_C^2-1}\sum_{a,b}\delta^{ab}q_{1_\perp}^\mu q_{2_\perp}^\nu V_{\mu\nu}^{ab} \; .
\end{equation}
Here $a$ and $b$ are colour indices, $M_X$ is the central object mass, $V_{\mu\nu}^{ab}$ represents the $gg \to X$ vertex and $q_{i_\perp}$ are the transverse momenta of the incoming gluons, given by
\begin{equation}
q_{1_\perp}=Q_\perp-p_{1_\perp}\,, \qquad
q_{2_\perp}=-Q_\perp-p_{2_\perp}\,,
\label{qperpdef}
\end{equation}
where $Q_\perp$ is the momentum transferred round the gluon loop while $p_{i_\perp}$ are the transverse momenta of the outgoing protons. Only one transverse momentum scale is taken into account in (\ref{bt}) by the prescription
\begin{align}\nonumber
Q_1 &= {\rm min} \{Q_\perp,|({\bf Q_\perp}-{\bf p}_{1_\perp})|\}\;,\\ \label{minpres}
Q_2 &= {\rm min} \{Q_\perp,|({\bf Q_\perp}+{\bf p}_{2_\perp})|\} \; .
\end{align}
The longitudinal momentum fractions carried by the gluons satisfy
\begin{equation}\label{xcomp}
\bigg(x' \sim \frac{Q_\perp}{\sqrt{s}}\bigg)  \ll \bigg(x \sim \frac{M_X}{\sqrt{s}}\bigg) \; ,
\end{equation} 
where $x'$ is the momentum fraction of the second $t$-channel gluon. The differential cross section at $X$ rapidity $y_X$ is then given by
\begin{equation}\label{ampnew}
\frac{{\rm d}\sigma}{{\rm d} y_X}=\langle S^2_{\rm enh}\rangle\int{\rm d}^2\mathbf{p}_{1_\perp} {\rm d}^2\mathbf{p}_{2_\perp} \frac{|T(\mathbf{p}_{1_\perp},\mathbf{p}_{2_\perp}))|^2}{16^2 \pi^5} S_{\rm eik}^2(\mathbf{p}_{1_\perp},\mathbf{p}_{2_\perp})\; ,
\end{equation}
where $T$ is given by (\ref{bt}) and $S^2_{\rm eik}$ is the `eikonal' survival factor, calculated using a generalisation of the `two-channel eikonal' model for the elastic $pp$ amplitude (see ~\cite{KMRtag} and references therein for details).

Besides the effect of eikonal screening $S_{\rm eik}$, there is some suppression caused by the rescatterings of the intermediate partons (inside the unintegrated gluon distribution, $f_g$). This effect is described by the so-called enhanced Reggeon diagrams and usually denoted as $S^2_{\rm enh}$, see Fig.~\ref{fig:pCp}. The value of $S^2_{\rm enh}$ depends mainly on the transverse momentum of the corresponding partons, that is on the argument $Q^2_i$ of $f_g(x,x',Q^2_i,\mu^2;t)$ in (\ref{bt}), and depends only weakly on the $p_\perp$ of the outgoing protons (which formally enters only at NLO). While $S^2_{\rm enh}$ was previously calculated (see~\cite{HarlandLang10,HarlandLang09}) using the formalism of~\cite{Ryskin09}, we now use a newer version of this model~\cite{Ryskin:2011qe} which includes the continuous dependence on $Q^2_i$ and not only three `Pomeron components' with different `mean' $Q_i$. We therefore include the $S_{\rm enh}$ factor inside the integral (\ref{bt}), with $\langle S^2_{\rm enh}\rangle$ being its average value integrated over $Q_\perp$.

If we consider the exact limit of forward outgoing protons, $p_{i_\perp}=0$, then we find that after the $Q_\perp$ integration (\ref{Vnorm}) reduces to~\cite{KMRprosp}
\begin{equation}\label{mprop}
\mathcal{M}\propto q_{1_\perp}^i q_{2_\perp}^j V_{ij} \to \frac{1}{2}Q_\perp^2(V_{++}+ V_{--})\sim\sum_{\lambda_1,\lambda_2}\delta^{\lambda_1\lambda_2}V_{\lambda_1\lambda_2}\;,
\end{equation}
where $\lambda_{(1,2)}$ are the gluon helicities in the $gg$ rest frame. The only contributing helicity amplitudes are therefore those for which the $gg$ system is in a $J_z=0$ state, where the $z$-axis is defined by the $gg$ direction which, up to corrections of order $\sim q_\perp^2/M_X^2$, is aligned with the beam axis.  In general, the outgoing protons can pick up a small $p_\perp$, but large values are strongly suppressed by the proton form factor, and so the production of states with non-$J_z=0$ quantum numbers is correspondingly suppressed (see~\cite{HarlandLang10,HarlandLang09} for an example of this in the case of $\chi_{(c,b)}$ and $\eta_{(c,b)}$ CEP). In particular, we find roughly that
\begin{equation}\label{simjz2}
\frac{|T(|J_z|=2)|^2}{|T(J_z=0)|^2} \sim \frac{\langle p_\perp^2 \rangle^2}{\langle Q_\perp^2\rangle^2}\;,
\end{equation}
which is of order $\sim1/50-1/100$, depending on the central object mass and cms energy $\sqrt{s}$. We shall see that this `$J_z=0$ selection rule'~\cite{Kaidalov03,Khoze:2000mw,Khoze00a} will have important consequences for the case of meson pair CEP\footnote{In the perturbative domain the hierarchy $p_\perp<Q_\perp$ is provided  by the mass inequality $M_X \gg M_{Y,Z}$, that is the mass of the centrally produced state, $M_X$, is large in comparison with the mass of the proton or, in the case of the process (\ref{dd}), the states, $M_Y,\ M_Z$, originating from the proton dissociation. 
Recall that normally in the hard process a large momentum transfer, $p_\perp$, (this is the momentum transferred through the $t$-channel $gg$ system, similar to that entering in (\ref{qperpdef}) for the purely exclusive process (\ref{exc})) leads to proton dissociation into higher mass states $Y$, $X$. Thus the probability to observe a system of low mass $M_{Y,Z}$ when there is a large $p_\perp$ transfer is small. The dominant contribution in the low mass, $M_{Y,Z}$, region comes from nucleon resonances where large values of $p_\perp$ are suppressed by the nucleon, or $N\to N^*$ transition, form factors. On the other hand, the Sudakov form factor hidden in the unintegrated gluon densities, $f_g$, strongly suppresses the contribution from low $Q_\perp \ll M_X$, leading to a rather large average $Q_\perp$ in the integral (\ref{bt}). Moreover, there are phenomenological indications (such as the small value of the slope, $\alpha'_P$, of the Pomeron trajectory, the success of the additive quark model, $\sigma(\pi p)\simeq (2/3)\sigma(pp)$, etc.) that the radius of the `soft' Pomeron is small and that it should have a qualitatively similar structure to the `hard' (QCD) Pomeron (see~\cite{Ryskin:2011qe,Ryskin:2009qf} for more details). Thus even for a relatively small $M_X$ we may consider the diagram shown in Fig.~\ref{fig:pCp} to have a relatively large average $Q_\perp$. Thanks to the fourth power $(p_\perp/Q_\perp)^4$ in (\ref{simjz2}), already for $Q_\perp\sim 2 p_\perp$ the $J_z=0$ selection rule becomes quite effective, with the $|J_z|=2$ contribution being suppressed by a factor of $\sim$ 16.}. Finally, we note that in (\ref{mprop}) the incoming gluon (transverse) helicities are averaged over at the {\it amplitude} level: this result is in complete contrast to a standard inclusive production process where the {\it cross section} is averaged over all gluon helicities. Eq. (\ref{mprop}) can be readily generalised to the case of non-$J_z=0$ gluons which occurs away from the forward proton limit, see in particular Section 4.1 (Eq. (41)) of~\cite{HarlandLang10}, which we will make use of throughout to calculate the $M\overline{M}$ CEP amplitudes from the corresponding $gg\to M\overline{M}$ helicity amplitudes.

\section{The $gg \to M\overline{M}$ perturbative process}

\subsection{Background: large angle $\gamma\gamma \to M\overline{M}$ meson pair production}\label{back}

Before considering the $gg\to M\overline{M}$ process relevant to CEP, we begin by summarising the case of $\gamma\gamma \to M\overline{M}$, which has been considered previously in the literature; the formalism used to describe this process can then be readily applied to the $gg$ case. The leading order contributions to $\gamma\gamma \to M\overline{M}$ were first calculated in~\cite{Brodsky81} (see also~\cite{Benayoun89,Chernyak06}), where $M(\overline{M})$ is a flavour nonsinglet meson(antimeson)\footnote{As pointed out in~\cite{Brodsky81}, see also~\cite{Atkinson83}, for the case of mesons with flavor-singlet Fock states there is also a contribution coming from the LO two-gluon component, see Section~\ref{eta}.}. They can be written in the form
\begin{equation}\label{amp}
\mathcal{M}_{\lambda\lambda'}(\hat{s},\theta)=\int_{0}^{1} \,{\rm d}x \,{\rm d}y\, \phi_M(x)\phi_{\overline{M}}(y)\, T_{\lambda\lambda'}(x,y;\hat{s},\theta)\;.
\end{equation}
where $\hat{s}$ is the $M\overline{M}$ invariant mass, $x,y$ are the meson momentum fractions carried by the quarks, $\lambda$, $\lambda'$ are the photon helicities and $\theta$ is the scattering angle in the $\gamma\gamma$ cms frame.
 $T_{\lambda\lambda'}$ is the hard scattering amplitude for the parton level process $\gamma\gamma\to q\overline{q}\,q\overline{q}$, where each (massless) $q\overline{q}$ pair is collinear and has the appropriate colour, spin, and flavour content projected out to form the parent meson. In the meson rest frame, the relative motion of the quark and antiquark is small: thus for a meson produced with large momentum, $|\vec{k}|$, we can neglect the transverse component of the quark momentum, $\vec{q}$, with respect to $\vec{k}$, and simply write $q=xk$ in the calculation of $T_{\lambda\lambda'}$. $\phi(x)$ is the meson wavefunction, representing the probability amplitude of finding a valence parton in the meson carrying a longitudinal momentum fraction $x$ of the meson's momentum, integrated up to the scale $Q$ over the quark transverse momentum $\vec{q_t}$ (with respect to pion momentum $\vec k$).

\begin{figure}
\begin{center}
\includegraphics[scale=1.2]{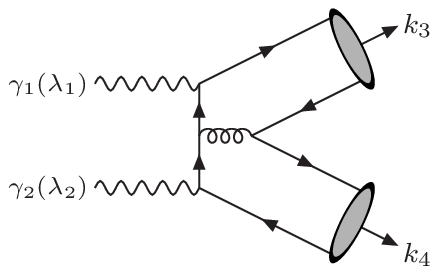}
\caption{A typical diagram for the $\gamma\gamma\to M\overline{M}$ process.}\label{gampi}
\end{center}
\end{figure}

A representative diagram contributing to $T_{\lambda\lambda'}$ is shown in Fig.~\ref{gampi}, where the additional hard gluon is required to supply the momentum transfer between the quark lines necessary for large angle scattering. There are four basic LO Feynman diagrams which contribute (the rest being given by simple permutations of the fermion lines and incoming photons), and after an explicit calculation we find\footnote{We have made use of the FORM symbolic manipulation programme throughout.}
\begin{align}\label{Tpp}
T^{++}_{\gamma\gamma}=T^{--}_{\gamma\gamma}&=-C_F\frac{4\pi\alpha_S}{\hat{s}}\frac{32 \pi \alpha}{x(1-x)y(1-y)}\bigg[ \frac{(e_1-e_2)^2 a}{1-\cos^2\theta}\bigg]\,\\ \nonumber
T^{+-}_{\gamma\gamma}=T^{+-}_{\gamma\gamma}&=C_F\frac{4\pi\alpha_S}{\hat{s}}\frac{32 \pi \alpha}{x(1-x)y(1-y)}\bigg[ \frac{(e_1-e_2)^2(a-1)}{1-\cos^2\theta}-\frac{e_1 e_2 a(y(1-y)+x(1-x))}{a^2-b^2\cos^2{\theta}}\\ \label{Tpm}
&-\frac{(e_1^2-e_2^2)(x-y)}{2}\bigg]\,
\end{align}
consistent with the results of~\cite{Brodsky81}, up to a universal `$-$' sign. $e_1$, $e_2$ are the quark charges (i.e. the mesons have charges $\pm (e_1-e_2)$) and
\begin{align}\label{a}
a&=(1-x)(1-y)+xy\; ,\\ \label{b}
b&=(1-x)(1-y)-xy\; .
\end{align}
A few comments are in order about the meson wavefunction $\phi(x)$. Firstly, as is usual for the factorisation of long- and short-distance physics, $\phi(x)$ must evolve with $Q^2$ in order for the physical predictions to remain factorisation scale invariant (see for example~\cite{Lepage80} for a detailed discussion). In particular, $\phi(x)=\phi(x,Q^2)$ obeys an evolution equation of the form
\begin{equation}\label{ev}
\frac{\partial\phi(x,Q)}{\partial\ln Q^2}=\frac{\alpha_S(Q^2)}{4\pi} \int_{0}^{1} \,{\rm d}y \,V(x,y)\,\phi(y,Q)\;,
\end{equation}
where $V$ can be computed from a single gluon exchange kernel~\cite{Lepage80}. While this specifies the evolution of $\phi$, the form it has at the starting scale $Q_0$ depends on the (non-perturbative) details of hadronic binding and cannot be predicted in perturbation theory. However, the overall normalisation is set by the meson decay constant $f_M$ via~\cite{Brodsky91}
\begin{equation}\label{wnorm}
\int_{0}^{1}\, {\rm d}x\,\phi_M(x,Q)=\frac{f_M}{2\sqrt{3}}\;.
\end{equation}
It was also shown in~\cite{Lepage80} that for very large $Q^2$ the meson wavefunction evolves towards the asymptotic form
\begin{equation}\label{asym}
\phi_M(x,Q)\underset{Q^2\to \infty}{\to} \,\sqrt{3} f_M\, x(1-x)\;.
\end{equation}
However the logarithmic evolution of (\ref{ev}) is very slow and at realistic $Q^2$ values the form of $\phi_M$ can in general be quite different. Indeed, the recent BABAR  data~\cite{Aubert09,Druzhinin09}, for example, strongly suggests that $\phi(x,Q)$ does not have the asymptotic form out to $Q^2\lesssim 40\,{\rm GeV}^2$. The issue of the correct form to take for $\phi(x)$ has been the subject of much theoretical interest (see for example~\cite{Bakulev01}--\cite{Chernyak09} and references therein), but we will not discuss this in any detail here. Instead, to calculate `benchmark' numerical results we will take $\phi(x,Q)$ to have the form proposed in~\cite{Chernyak81}
\begin{equation}\label{CZ}
\phi^{{\rm CZ}}(x,Q_0)=5\sqrt{3}f_M\, x(1-x)(2x-1)^2\;.
\end{equation}
It is shown in for example~\cite{Chernyak09} that the available $\gamma\gamma\to M\overline{M}$ data are quite well described by this choice, while this is in general not the case for the asymptotic form (\ref{asym}). For simplicity we do not include the evolution of (\ref{CZ}) in our calculation:\footnote{It is well known that the annihilation of two energetic (colour charged) gluons into a colourless system (two mesons in this case) will in general be accompanied by an intensive bremsstrahlung of relatively soft gluons. An {\it exclusive} cross section is suppressed by the small probability of not having such radiation, which is given by the Sudakov form factor, $T$, incorporated in the unintegrated gluon densities, $f_g$, in (\ref{bt})-- see~\cite{Khoze00,HarlandLang09} for more details. However, there may also be analogous radiation from the quarks in the final state, when the small ($\sim 1/Q$) size $q\bar q$ dipole forms a pion of the `normal' ($\sim 0.6$ fm) size. The corresponding form factor should be accounted for in the evolution (\ref{ev}) of the meson wavefunction and in this way included in the value of $\phi(x,Q)$. Moreover, if we choose a relatively low scale $Q^2 \ll s_{\pi\pi}$ then we have to multiply the result by an additional Sudakov form factor, $T(Q^2,s_{\pi\pi})$, describing the probability not to emit other gluons with momenta from $Q$ to $\sqrt s_{\pi\pi}$. Thus we may justify the use of the phenomenological form (\ref{CZ}) only in a limited interval of $s_{\pi\pi}\sim 10 -20$ GeV$^2$. For a larger pion pair subenergy we may expect an additional Sudakov-like suppression.}
 for the realistic transverse momenta, $k_\perp$, values considered in this paper, the scale $Q\sim {\rm min}(x,1-x)k_\perp$ is sufficiently low that the effect of including the evolution of $\phi(x)$ is very small, and well within other theoretical uncertainties, coming from in particular the choice of $\phi(x)$ at the starting scale, $Q_0$.

Finally, returning to the hard amplitudes (\ref{Tpp}) and (\ref{Tpm}), a few comments are in order. Firstly, we can see that the amplitudes are divergent as $\cos\theta\to\pm 1$, that is when the incoming photons and quark lines become collinear. For small angle scattering, where there is no longer a large momentum transferred between the incoming photons and the quark lines, the quark propagators become soft and fixed-order perturbation theory can no longer be trusted. As we will see in Section~\ref{eta}, upon the inclusion of finite quark masses, $m_q$, the amplitudes  are finite at $\cos\theta=\pm 1$, although clearly LO perturbation theory in this region can still not be trusted. In practice a reasonable cut-off is simply imposed on $|\cos\theta|$ to avoid this non-perturbative region of phase space: experimentally, for $M\overline{M}$ CEP this simply corresponds to demanding that the mesons are produced relatively centrally, and indeed in reality this is always the case.

Secondly, we can see in the case of neutral meson ($e_1=e_2$), for example $\pi^0\pi^0$, production that the $T^{++}_{\gamma\gamma}$, $T^{--}_{\gamma\gamma}$ amplitudes (when the incoming  photons are in a $J_z=0$ state along the $\gamma\gamma$ axis) vanish; we will see in Section \ref{bg} that a similar result holds for the $gg\to\pi\pi$ process. For charged pion $\pi^+\pi^-$ production both the $|J_z|=2$ and $J_z=0$ amplitudes contribute, while in the $\pi^0\pi^0$ case only one term in the $|J_z|=2$ amplitudes does, and so we would expect the $\pi^0\pi^0$ cross section to be suppressed, with an explicit calculation, using (\ref{CZ}) for the pion wavefunction, giving $\sigma(\pi^0\pi^0)/\sigma(\pi^+\pi^-)\approx 0.03$ for $|\cos\theta|<0.6$ (see also Fig. 3 of~\cite{Brodsky81}). However, it should be noted that recently BELLE~\cite{:2009cka} have reported the significantly large value of $\sigma(\pi^0\pi^0)/\sigma(\pi^+\pi^-)=0.32\pm0.03\pm0.05$ in the range $\sqrt{\hat{s}}=3.1-4.1$ GeV and $|\cos\theta|<0.6$. We would argue that the $\hat{s}$ values being probed may be too small to justify the leading-twist pQCD approximation outlined above (see also~\cite{Chernyak09} for some discussion of this) in the $\pi^0\pi^0$ case where the formally leading amplitudes are strongly suppressed. For example, if we allow for the fact that the $q\overline{q}$ pair that form the pion can have a non-zero  $q_t$ then the exact cancellation in (\ref{Tpp}) will not in general occur, and this could provide some enhancement to $\sigma(\pi^0\pi^0)/\sigma(\pi^+\pi^-)$. 

\subsection{$gg \to M\overline{M}$: Feynman diagram calculation}\label{bg}

\begin{figure}[h]
\begin{center}
\includegraphics[scale=0.6]{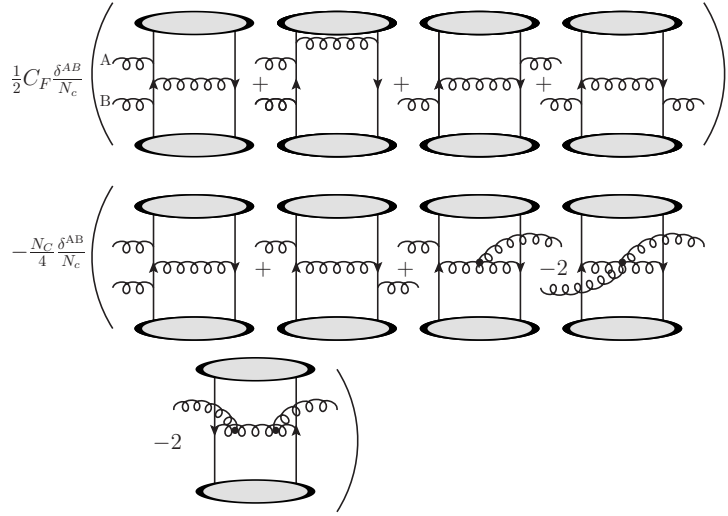}
\caption{Basic Feynman diagrams for the $gg\to M\overline{M}$ process, grouped into individually gauge invariant subsets $T_1$ (upper) and $T_2$ (lower), and with the relevant colour factors shown schematically. The inclusion of all permutations of these diagrams is implicit.}\label{gg}
\end{center}
\end{figure}
The $gg \to M\overline{M}$ process can be calculated in analogy to the $\gamma\gamma$ reaction, that is using the formalism of (\ref{amp}), but where the $gg \to q\overline{q}\,q\overline{q}$ parton-level amplitudes are to be evaluated. There are now seven independent Feynman diagrams: four `abelian' diagrams, where the photons are simply interchanged with gluons, and three additional `non-abelian' diagrams, where three and four-gluon vertices are present. For ease of calculation we can group these into two separate, individually gauge invariant, sets of diagrams, each weighted with the same colour factor, as shown in Fig.~\ref{gg}. In particular, we can decompose the $T_{gg}$ amplitude as
\begin{equation}
T_{gg}= \frac{\delta^{\rm AB}}{N_C}\bigg(\frac{1}{2}C_F T_1-\frac{N_C}{4}T_2\bigg)\; ,
\end{equation}
where $T_1$ and $T_2$ represent the corresponding amplitudes for the diagram sets in Fig.~\ref{gg}. $T_1$ is given by the kinematic part (i.e. with the colour factor $C_F\to 1$) of the $\gamma\gamma \to M\overline{M}$ amplitudes, (\ref{Tpp})-(\ref{Tpm}), by setting $e_1=e_2=1$ and $\alpha \to \alpha_S$. As we have $e_1=e_2$, the $J_z=0$ amplitudes for this vanishes (as in the $\gamma\gamma\to \pi^0\pi^0$ case), while for the $|J_z|=2$ amplitudes we find
\begin{equation}
T_1^{+-}=T_1^{-+}=\frac{1}{\hat{s}xy(1-x)(1-y)}\frac{-128 \pi^2\alpha_S^2 (a-b^2)a}{a^2-b^2\cos^2{\theta}}\; ,
\end{equation}
where $a,b$ are defined in (\ref{a})-(\ref{b}). An explicit calculation of the $T_2$ helicity amplitudes then gives
\begin{align}
T_2^{++}=T_2^{--}&=0\; ,\\
T_2^{+-}=T_2^{-+}&=\frac{1}{\hat{s}xy(1-x)(1-y)}\frac{-128\pi^2 \alpha_S^2 (a-b^2)\cos^2{\theta}}{a^2-b^2\cos^2{\theta}}\;,
\end{align}
giving for the total $gg$ amplitudes
\begin{align}\label{T++}
T_{gg}^{++}=T_{gg}^{--}&=0\;,\\ \label{T+-}
T_{gg}^{+-}=T_{gg}^{-+}&=\frac{\delta^{\rm AB}}{N_C}\frac{64\pi^2\alpha_S^2}{\hat{s}xy(1-x)(1-y)}\frac{(x(1-x)+y(1-y))}{a^2-b^2\cos^2{\theta}}\frac{N_C}{2}\bigg(\cos^2{\theta}-\frac{2 C_F}{N_C}a\bigg)\;,
\end{align}
the gauge invariance of which has been confirmed by explicit calculation.

We have therefore shown that the $gg\to M\overline{M}$ amplitude for $J_z=0$ gluons vanishes at LO for flavour-nonsinglet mesons, which, recalling the $J_z=0$ selection rule that strongly suppresses the CEP of non-$J_z=0$ states, will lead to a strong suppression in the $M\overline{M}$ production cross section, see (\ref{simjz2}). However it should be noted that any NNLO corrections\footnote{As there is no contribution from the interference between the $O(\alpha_S^3)$ 1-loop and the vanishing $O(\alpha_S^2)$ tree level $J_z=0$ amplitudes, there are no NLO corrections in CEP, and the first non-zero perturbative correction to the cross section must enter at NNLO.}, or higher twist effects\footnote{In particular, corrections that result from allowing for a non-zero quark $q_t$ with respect to the meson momentum, expected to be roughly of order the constituent quark mass $q_t \sim 350$ MeV, may be important for the lower values of $M_X$ that are experimentally relevant.} which allow a $J_z=0$ contribution may cause the precise value of the cross section to be somewhat larger than the leading-order, leading-twist $|J_z|=2$ estimate, although qualitatively the strong suppression will remain. This is in particular true when, as we shall discuss below, the $|J_z|=2$ amplitude is additionally suppressed by the presence of a radiation zero in the amplitude.

\begin{figure}[h]
\begin{center}
\includegraphics[scale=0.65]{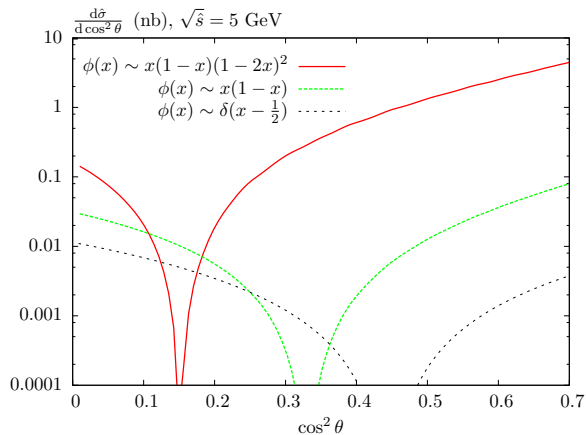}
\caption{Differential cross section ${\rm d}\sigma/{\rm d}|\cos \theta|$ at $\sqrt{\hat{s}}=5$ GeV, for the $gg\to M\overline{M}$ process for non-favour singlet scalar mesons. For comparison, the distribution for three choices of meson wavefunction are shown, the asymptotic form $\phi(x)\propto x(1-x)$, the form (\ref{CZ}) proposed in~\cite{Chernyak81}, and a $\delta$-function $\phi(x)\propto\delta(x-\frac{1}{2})$.}\label{rad0}
\end{center}
\end{figure}

We can see that the $|J_z|=2$ amplitude (\ref{T+-}) vanishes for a particular value of $\cos^2\theta$. This behaviour, which at first sight may appear quite unusual, is in fact not completely unexpected: the vanishing of a Born amplitude for the radiation of massless gauge bosons, for a certain configuration of the final state particles is a known effect, usually labelled a `radiation zero', see for instance~\cite{Brodsky0s82,Zhu80,Heyssler:1997ng,Brown95} and references therein. It results from the complete destructive interference of the classical radiation patterns, leading to a vanishing of the amplitude, and the general conditions for the existence of these zeros (which often do not occur in the physical phase space region) are derived in detail in~\cite{Brown82}. The position of the zero is determined by an interplay of both the internal (in the present case, colour) and space-time (the particle $4$-momenta) variables, as can be seen in (\ref{T+-}), where the position of the zero depends on the choice of meson wavefunction, $\phi(x)$, through the variables $a$ and $b$, as well as on the QCD colour factors, see Fig.~\ref{rad0}. In particular, the zero occurs in (\ref{T+-}) when $\cos^2\theta \approx 2C_F\langle a \rangle/N_C$, where $\langle a \rangle$ is the average value of $a$ integrated over the meson wavefunctions $\phi(x),\phi(y)$. As $0\leq a \leq 1$ for all physical values of $x,y$ and $2 C_F/N_C<1$ for all $N_C$ (while the prefactors in (\ref{T+-}) are strictly positive), this will always occur in the physical region for any SU(N) gauge theory.

While this effect, which is present in all theories with massless gauge bosons, is expected to occur in QCD, it is usually neutralised along with colour by the averaging of hadronisation. The CEP process, for which the fusing gluons are selected to be in a colour singlet state by the exclusivity of the event, is therefore in principle uniquely positioned to observe these zeros. However, it should again be noted that, as discussed above, as the $|J_z|=2$ amplitude is strongly suppressed by the $J_z=0$ selection rule, any NNLO or higher twist effects which allow a $J_z=0$ component to the cross section may give comparable contributions; it is therefore not clear that such a zero would in this case be seen clearly in the data (at lower values of $M\overline{M}$ invariant mass, where a non-perturbative framework is applicable, we would also not expect to find such a zero, see Section~\ref{nsect}). On the other hand, the destructive interference effects which lead to the zero in the $|J_z|=2$ amplitude (\ref{T+-}) will tend to suppress the CEP rate.

\subsection{$gg\to M\overline{M}$: MHV calculation}

It is well known (see for example~\cite{Mangano90}) that the tree level $n$-gluon scattering amplitudes in which the maximal number ($n-2$) of gluons have the same helicity, the so-called `maximally helicity violating' (MHV), or `Parke-Taylor', amplitudes, are given by remarkably simple formulae~\cite{Parke86,Berends87}. These results were extended using supersymmetric Ward identities to include amplitudes with one and two quark-antiquark pairs in~\cite{Mangano90,Mangano87,Mangano88}, where `MHV' refers to the case where ($n-2$) partons have the same helicity. In these cases, simple analytic expressions can again be written down for the MHV amplitudes, while for greater than 2 fermion-antifermion pairs (recalling that the helicities of a connected fermion-antifermion pair must be opposite) no MHV amplitudes exist. More recently, it has been shown in~\cite{Cachazo04} that the $n$-gluon scattering amplitude for {\it any} helicity configuration can be calculated with this formalism; in particular they can be constructed from tree graphs in which the vertices are the usual tree-level MHV scattering amplitudes continued off-shell in a specific way. This was extended to include fermions in~\cite{Georgiou04}.

The basis of all of these results lies in the spinor helicity formalism (see~\cite{Mangano90,Dixon96} for reviews): one calculates the matrix elements with the external states having a given assigned helicity as an expression written in terms of spinor products of the external particle momenta. These amplitudes can then be evaluated, usually numerically, squared and summed incoherently to give the full cross section. Some basic formulae are given in Appendix~\ref{mhvf}.

As a further check of (\ref{T++}) and (\ref{T+-}) we can therefore calculate the $gg\to q\overline{q}q\overline{q}$ helicity amplitudes using the spinor helicity formalism described above. For the $J_z=0$ case, the amplitudes ($T_{++}$ and $T_{--}$) are MHV and, as we shall show, the vanishing of these amplitudes follows simply from the known Parke-Taylor amplitudes. We will therefore consider these first, as it is simpler than the $|J_z|=2$ case, and will serve as a clearer example.
\subsubsection{$J_z=0$ amplitudes}
In general it is well known that the full $n$-parton amplitude $\mathcal{M}_n$ can be written in the form of a `dual expansion', as a sum of products of colour factors $T_n$ and purely kinematic partial amplitudes $A_n$
\begin{equation}\label{mhv}
\mathcal{M}_n(\{p_i,h_i,c_i\})=\sum_\sigma T_n(\{c_{\sigma(i)}\})A_n(\{k_{\sigma(i)},h_{\sigma(i)}\})\;,
\end{equation}
where $\{c_{\sigma(i)}\}$ are colour labels and $\{k_{\sigma(i)},h_{\sigma(i)}\}$ are the momenta and helicities, respectively, of the external legs $i=1\cdots n$, and the sum is over appropriate simultaneous non-cyclic permutations $\sigma$ of colour labels and kinematics variables. The colour factors $T_n$ are easy to determine, while the purely kinematic part of the amplitude $A_n$ is to be calculated, and encodes all the non-trivial information about the full amplitude, $\mathcal{M}_n$, see for instance~\cite{Georgiou04} for more details.

We are therefore interested in calculating the kinematic amplitudes for the 6-parton $g(\pm)g(\pm)\to q\overline{q}q\overline{q}$ process: once this is done we simply use (\ref{mhv}) to determine the full amplitudes. These MHV amplitudes have very simple forms (given in full in~\cite{Birthwright05} and elsewhere), and in fact the total $n$-point amplitude for $q\overline{q}q\overline{q}$ plus $(n-4)$ positive helicity gluons can be written down in two lines~\cite{Mangano90}
\begin{align}\nonumber
M_n&=i g^{n+2} A_0(h_s, h_r, h_g) \sum_\sigma \frac{\langle k_s\,k_{\overline{r}}\rangle}{\langle k_s\,a_1\rangle \cdots \langle a_l\,k_{\overline{r}}\rangle}\frac{\langle k_r\,k_{\overline{s}}\rangle}{\langle k_r\,b_1\rangle \cdots \langle b_{l'}\,k_{\overline{s}}\rangle}(\lambda^{a_1} \cdots \lambda^{a_l})_{i_1 j_2}(\lambda^{b_1} \cdots \lambda^{b_{l'}})_{i_2 j_1}\\ \label{PT}
&-\frac{1}{N_C}\frac{\langle k_s\,k_{\overline{s}}\rangle}{\langle k_s\,a_1\rangle \cdots \langle a_l\,k_{\overline{s}}\rangle}\frac{\langle k_r\,k_{\overline{r}}\rangle}{\langle k_r\,b_1\rangle \cdots \langle b_{l'}\,k_{\overline{r}}\rangle}(\lambda^{a_1} \cdots \lambda^{a_l})_{i_1 j_1}(\lambda^{b_1} \cdots \lambda^{b_{l'}})_{i_2 j_2}\;.
\end{align}
Here the indices $r(\overline{r})$ and $s(\overline{s})$ refer to the quarks (antiquarks) with colour indices $i_1(j_1)$ and $i_2(j_2)$, respectively, and the labels $a_i$, $b_i$ refer to the gluons, while the standard spinor contraction `$\langle k,l\rangle$' is defined in (\ref{spina}). The sum is over all the partitions of the gluons ($l+l'=n-4$, $l=0,\cdots,n-4$) and over the permutations of the gluon indices, with the product of zero $\lambda$ matrices becoming a Kronecker delta and the kinematical factors equal to one when $l=(0,n-4)$. The overall factor $A_0$ depends on the particular quark helicity configuration: for our calculation, it is given by the expressions from~\cite{Mangano90} for the two quark helicity combinations relevant to the meson spin projections. For the equivalent `$\overline{{\rm MHV}}$' diagram with all negative helicity gluons, we simply replace $\langle k\,l\rangle \to [k\,l]$, see (\ref{spinb}).

Considering now the case of the 6-parton amplitude relevant to our calculation, we make the following identifications
\begin{align}\label{qs}
k_r&=x k_3\quad k_{\overline{r}}=(1-y) k_4 \quad k_s=y k_4 \quad k_{\overline{s}}=(1-x)k_3\;,\\
i_1&=j_2 \quad i_2=j_1\;,
\end{align}
for the collinear quarks (neglecting as usual the $q_t$ of the quarks relative to the meson momenta) to form colour singlet mesons, where $k_{3,4}$ are defined as in Fig.~\ref{gampi}. Note that there is in general a second possible assignment corresponding to the diagrams for which the $r\overline{r}$ and $s\overline{s}$ pairs belong to the same mesons, but this does not contribute for non-isosinglet states; we shall discuss this further in Section~\ref{eta}.

Immediately we can see that first term in (\ref{PT}) goes like $\sim k_3^2, k_4^2=0$, while the colour factors for the individual pieces contributing to the second term are universal and are given by ${\rm Tr}(\lambda^a \lambda^b)=\delta^{ab}/2$. Factoring this out, we readily find that the amplitude is given by
\begin{align}\nonumber
M&\propto \frac{\langle k_3\, k_4 \rangle}{ \langle k_4\, k_1 \rangle \langle k_1\, k_3 \rangle \langle k_3\, k_2 \rangle \langle k_2\, k_4 \rangle}+\frac{1}{\langle k_3\, k_1 \rangle\langle k_1\, k_2 \rangle
\langle k_2\, k_4 \rangle}+\frac{1}{\langle k_3\, k_2 \rangle \langle k_2\, k_1 \rangle \langle k_1\, k_4 \rangle}\\ \label{canc}
&\propto\langle k_3\, k_2 \rangle\langle k_1\, k_4 \rangle+\langle k_1\, k_3 \rangle\langle k_2\, k_4 \rangle-\langle k_3\, k_4 \rangle\langle k_1\, k_2 \rangle=0\;,
\end{align}
from the Schouten identity (\ref{sc}), while the $\overline{{\rm MHV}}$ amplitude similarly vanishes. This result depends crucially on the colour structure and collinearity of the (massless) quarks/antiquarks given in (\ref{qs}), which lead to the factorisation of the colour factors and the cancellation between the kinematic pieces in (\ref{canc}), respectively. It also requires 
that the produced mesons are flavour non-singlet states, see Section~\ref{eta} for a discussion of this. 

We have therefore in a few lines of algebra confirmed the vanishing of the $gg \to M\overline{M}$ amplitudes  for $J_z=0$ initial-state gluons (\ref{T++}), which resulted from a non-trivial calculation of 7 independent Feynman diagrams. This gives some idea of the power of the MHV formalism, which we now apply to the more complicated non-MHV $|J_z|=2$ case.

\subsubsection{$|J_z=2|$ amplitudes}

\begin{figure}[h]
\begin{center}
\includegraphics[scale=1]{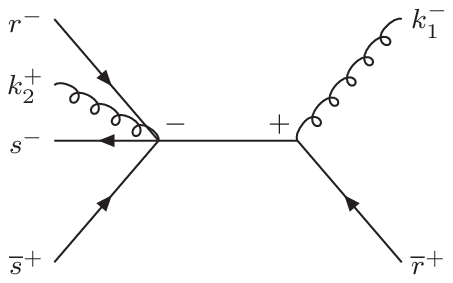}
\includegraphics[scale=1]{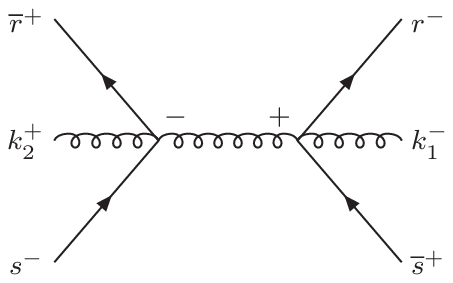}
\includegraphics[scale=1]{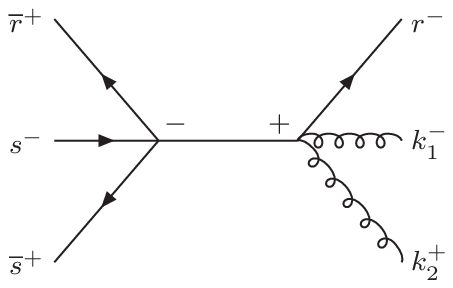}
\caption{Representative tree diagrams contributing to the $g(k_1)g(k_2)\to q\overline{q}q\overline{q}$ process with $|J_z|=2$ incoming gluons. Quark labels follow the same notation as (\ref{PT}) and $\pm$ signs represent particle helicity, with all momenta defined as incoming. All contributing amplitudes are of these three types.}\label{jz2}
\end{center}
\end{figure}

To calculate these amplitudes, which are not MHV, we follow the formalism described in~\cite{Georgiou04,Wu04}. The basic idea is that these `nMHV' diagrams can be calculated by connecting two MHV diagrams in all allowed ways with a scalar propagator $1/p^2$. For the general $q\overline{q}q\overline{q}$ case there are four types of diagram that contribute (shown in Fig. 2 of~\cite{Georgiou04}). In fact the set of diagrams with a purely gluonic MHV sub-graph can readily be shown to vanish when the colour singlet projection is performed and all permutations are summed over (in particular, the amplitudes for the diagrams where the external gluon legs are interchanged have a relative minus sign but are otherwise identical), and so we will not consider these further. Representative diagrams for the three contributing sets are shown in Fig.~\ref{jz2}. The remaining diagrams are then given by including all possible positions and permutations of the two gluons, with the important requirement that gluons are always emitted from the same side of the connected quark-antiquark line (see in particular Fig. 1 of ~\cite{Wu04} and the discussion in the text).

As described above, this calculation method relies on the factorisation of the total amplitude $\mathcal{M}_n$ into the sum of products of kinematic partial amplitudes $A_n$ and colour factors $T_n$. The rules for calculating the kinematic amplitudes are explained in~\cite{Georgiou04}, while the colour factors are given in~\cite{Mangano90,Georgiou04}. For the specific case of the $N=6$ parton $gg \to q\overline{q}q\overline{q}$ amplitude, the colour factors are given by
\begin{equation}
T_n=\frac{(-1)^p}{N_C^p}(\lambda^{a_1} \cdots \lambda^{a_l})_{i_1 \alpha_1}(\lambda^{b_1} \cdots \lambda^{b_{l'}})_{i_2 \alpha_2}
\end{equation}
where $i_1,i_2$ are the colour indices of the quarks, $\alpha_1,\alpha_2$ are the colour indices of the antiquarks and the labels $a_i$, $b_i$ refer to the gluons. The pair $\{\alpha\}=(\alpha_1 \alpha_2)$ is a permutation of the pair $\{j\}=(j_1 j_2)$, where quark $i_k$ is connected by a fermion line to antiquark $j_k$. $p$ is then the number of times $\alpha_k=j_k$, with $p=1$ if $\{\alpha\}\equiv\{j\}$. As in (\ref{PT}), the sum is over all the partitions of the gluons ($l+l'=n-4$, $l=0,\cdots,n-4$) and over the permutations of the gluon indices, with the product of zero $\lambda$ matrices becoming a Kronecker delta. For a given $T_n$, the corresponding diagram has the same cyclic ordering of the quark and gluons as their colour labels in $T_n$.

For the two gluon case we can have $p=0,1$ and therefore expect two separate sets of diagrams with different colour coefficients to contribute, as can be seen in (\ref{PT}) for the $J_z=0$ amplitude. In particular, recalling the colour singlet assignment of (\ref{qs}), the only non-zero colour factors are of the form
\begin{align}
p&=0:\qquad(\lambda^{a_1}\lambda^{a_2})_{i_1 j_2}\delta_{i_2 j_1} \to \frac{N_C}{2}\delta^{a_1 a_2}\;,\\
p&=1:\qquad-\frac{1}{N_C}(\lambda^{a_1}\cdots)_{i_1 j_1}(\lambda^{a_2}\cdots)_{i_2 j_2} \to -\frac{1}{2 N_C}\delta^{a_1 a_2}\;.
\end{align}
The diagrams shown in Fig.~\ref{jz2} correspond to $p=1$: the $p=0$ diagrams are given by making the replacement $\overline{r}\leftrightarrow \overline{s}$, with the important requirement that a $q\overline{q}$ pair of the same flavour must be connected by a fermion line. We can decompose the previous result for the $|J_z|=2$ amplitude (\ref{T+-}) in a similar way
\begin{equation}
T_{gg}^{+-}=T_{gg}^{-+}\propto (N_C(a-\cos^2 \theta)-\frac{1}{N_C}a)\;.
\end{equation}
Following a fairly lengthy explicit calculation, we have then showed numerically that the sum of the $p=0,1$ amplitudes reproduce these two colour-decomposed terms, thus confirming the result of (\ref{T+-}).

We end this section with some final words of explanation. Firstly, to the diagrams shown in Fig.~\ref{jz2} we must also add the diagrams where the quarks have positive helicity (the helicity of the antiquark must then be negative), corresponding to the second term in the meson spin-0 projection; these are obtained by simply interchanging $x\leftrightarrow y$. Secondly, in implementing the off-shell prescription required to join the MHV sub-graphs, we must choose a reference spinor with which to define certain spinor products, see~\cite{Georgiou04} for more details. In our calculation we take this to be defined by the gluon momentum $|p_1\rangle$. While each individual amplitude depends on this choice, of course when the amplitudes are summed this dependence must cancel. However, for certain diagrams unphysical infinities (that is, unrelated to standard IR soft/collinear divergences) can arise when external momenta are collinear to the reference momentum which, while cancelling in the total amplitude, must be dealt with carefully (see, for example~\cite{Kleiss85}). This is achieved by making the replacement $|p_1\rangle\to|p_1\rangle+|\epsilon\rangle$, bringing the divergent amplitudes to a common denominator and then using Schouten's identity (\ref{sc}) to simplify the resulting expression, before setting $|\epsilon\rangle$ to zero, at which stage the result will be finite. Finally, we note that to correctly calculate the $gg\to M\overline{M}$ amplitudes for nonsinglet mesons, care must be taken to omit the set of diagrams which only contribute for flavour singlet states, see Section~\ref{eta}. These correspond to interchanging the quark legs in the Feynman diagrams in Fig.~\ref{gg}, so that the collinear $q\overline{q}$ pairs forming the mesons are connected by a fermion line, see Fig.~\ref{ladder}. We can therefore omit these diagrams by requiring that the two $q\overline{q}$ pairs in the $gg\to q\overline{q} q\overline{q}$ process are distinguishable, and using the relevant MHV rules for this case (these are given in~\cite{Birthwright05}, for example). Using the MHV rules for identical fermions (which includes the permutation between the fermion pairs) will not give the required result, as it will implicitly include the flavour singlet contribution.

\subsection{$gg \to V\overline{V}$ amplitudes}\label{svect}

The amplitudes $T^{gg}_{\lambda_1\lambda_2,\lambda_3\lambda_4}$ for the $g(\lambda_1)g(\lambda_2)\to V(\lambda_3)\overline{V}(\lambda_4)$ process, where $V(\overline{V})$ are helicity $\pm 1$ spin-1 mesons can readily be calculated using the formalism of Sections \ref{back} and \ref{bg}, and are for completeness given here. As we will see in Section~\ref{eta}, for vector meson production only diagrams of the type shown in Fig.~\ref{gg} contribute: in this case, for the helicity-0 state, the amplitudes are identical to those for scalar mesons -- see (\ref{T++}) and (\ref{T+-}). We find
\begin{align}\label{Tv1}
T^{gg}_{++,+-}&=T^{gg}_{++,-+}=T^{gg}_{--,+-}=T^{gg}_{--,-+}=0\;.\\\label{Tv2}
T^{gg}_{+-,+-}&=T^{gg}_{-+,-+}=-\frac{\delta^{ab}}{N_C}\frac{64\pi^2\alpha_S^2}{\hat{s}xy(1-x)(1-y)}\bigg(C_F b^2-\frac{N_C}{2}a\bigg)\frac{\cos \theta (1+\cos \theta)}{a^2-b^2 \cos^2 \theta}\;,\\\label{Tv3}
T^{gg}_{-+,+-}&=T^{gg}_{+-,-+}=\frac{\delta^{ab}}{N_C}\frac{64\pi^2\alpha_S^2}{\hat{s}xy(1-x)(1-y)}\bigg(C_F b^2-\frac{N_C}{2}a\bigg)\frac{\cos \theta (1-\cos \theta)}{a^2-b^2 \cos^2 \theta}\;,
\end{align}
where it is clear (in the limit that the quark $q_t=0$) that $T^{gg}_{\lambda_1\lambda_2,++}=T^{gg}_{\lambda_1\lambda_2,--}=0$, as helicity must be conserved along the fermion line for massless quarks. This result immediately follows from the helicity conserving gluon-$q\overline{q}$ vertices which enter the perturbative calculation, and the fact that the meson helicity is given by the sum of the helicities of its valence quarks: this forms the basis of the so-called `hadronic helicity conservation' selection rule, see~\cite{Brodsky81h}. The vanishing of the $J_z=0$ amplitudes (\ref{Tv1}) also occurs in the $\gamma\gamma\to V\overline{V}$ amplitude, and follows from (\ref{PT}) in exactly the same way as for the $gg \to M\overline{M}$ case\footnote{In fact, in the case of the $gg\to V\overline{V}$ amplitude, the prefactor $A_0$ in (\ref{Tv1}) also vanishes.}. The $|J_z|=2$ amplitudes could in principle be derived using the MHV formalism, although we do not consider that here.

\section{${\rm SU}(3)_F$ singlet contribution: $gg \to \eta\eta,\eta'\eta',\eta\eta'$}\label{eta}
\begin{figure}[h]
\begin{center}
\subfigure[]{\includegraphics[clip,trim=-20 0 0 0,scale=0.8]{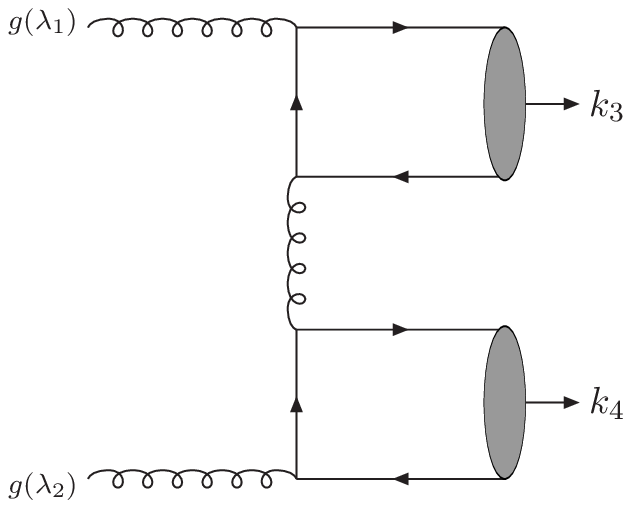}}\qquad
\subfigure[]{\includegraphics[clip,trim=-20 0 0 0,scale=0.8]{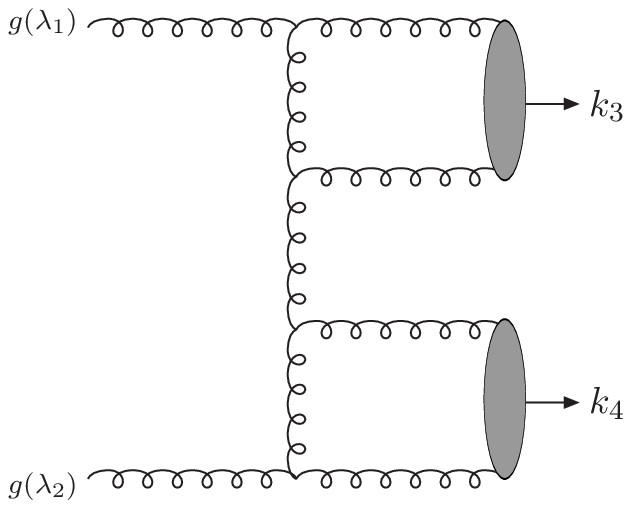}}
\caption{Representative `ladder' diagrams, which contribute to $gg\to M\overline{M}$ production for flavour-singlet mesons. (a) Valence $q\overline{q}$ contribution. (b) Valence $gg$ contribution.}\label{ladder}
\end{center}
\end{figure}
In general, as well as the diagrams shown in Fig. \ref{gg}, it is also possible for the $q\overline{q}q\overline{q}$ final state to form collinear colour singlet meson states via the process shown in Fig. \ref{ladder} (a). The contribution from this individually gauge invariant subset of `ladder' diagrams could in principle spoil the $J_z=0$ cancellation of (\ref{T++}). However, for the production of isovector states such as pions these diagrams violate isospin conservation and will therefore vanish: for example, for $\pi^0\pi^0$ production the $|u\overline{u}\rangle$ and $|d\overline{d}\rangle$ contributions interfere destructively to give zero total contribution.

Calculating the relevant amplitudes in the usual way we find for scalar mesons, for the valence quark contribution (these results could also be derived using the MHV formalism described previously, although we do not consider it here)
\begin{align}\label{lad0}
T_{++}^{\rm lad.}=T_{--}^{\rm lad.}&=\frac{\delta^{ab}}{N_C}\frac{64\pi^2 \alpha_S^2}{\hat{s}xy(1-x)(1-y)}\frac{(1+\cos^2 \theta)}{(1-\cos^2 \theta)^2}\;,\\ \label{lad2}
T_{+-}^{\rm lad.}=T_{-+}^{\rm lad.}&=\frac{\delta^{ab}}{N_C}\frac{64\pi^2 \alpha_S^2}{\hat{s}xy(1-x)(1-y)}\frac{(1+3\cos^2 \theta)}{2(1-\cos^2 \theta)^2}\;.
\end{align}
While these amplitudes do not contribute, for example, for $\pi\pi$ production, they will in general be relevant for $\eta'\eta'$ production and, through $\eta-\eta'$ mixing, $\eta\eta$ production. The ${\rm SU}(3)_F$ basis states are defined as
\begin{align}\label{eta0}
|\eta_0\rangle &= \frac{1}{\sqrt{3}}|u\overline{u}+d\overline{d}+s\overline{s}\rangle\;,\\ \label{eta8}
|\eta_8\rangle &= \frac{1}{\sqrt{6}}|u\overline{u}+d\overline{d}-2s\overline{s}\rangle\;.
\end{align}
In terms of these states the $\eta$ and $\eta'$ wavefunctions are given by
\begin{align}
|\eta\rangle&=\cos \theta_P |\eta_8\rangle - \sin \theta_P |\eta_0\rangle\;,\\
|\eta'\rangle&=\sin \theta_P |\eta_8\rangle + \cos \theta_P |\eta_0\rangle\;.
\end{align}
In the case of $\eta\eta$ CEP, the flavour singlet state $|\eta_0\rangle$ receives a non-zero $J_z=0$ contribution through (\ref{lad0}), and may in general greatly enhance the cross section relative to $\pi\pi$ production. There is some variation in the values of $\theta_P$ extracted from data, depending on the theoretical assumptions used in the analysis: following chiral perturbation theory, for example, the meson masses in the $SU(3)_F$ nonet can be used to extract a value $\theta_P=11.5^\circ$~\cite{Nakamura10}, but from fits to the decays of different light vector and pseudoscalar mesons (see for instance~\cite{Thomas07,Feldmann98} and references therein), a value of $\theta_P\approx -15^\circ$ is favoured, and we will take this value as a default\footnote{Often in the literature, rather than using the singlet-octet basis given in (\ref{eta0}) and (\ref{eta8}), a `quark' basis is taken, with corresponding mixing angle $\phi$. This is related to $\theta_P$ via $\theta_P=\phi-\arctan \sqrt{2}$.}. In this case, the flavour singlet contribution will be suppressed by a factor $\sin^4 \theta_P \sim 1/200$, which may therefore be comparable to the $|J_z|=2$ flavour-octet contribution. Moreover, the flavour-singlet amplitude is further enhanced by the larger multiplicity of flavour states that can contribute to the independent $q\overline{q}$ pairs in Fig.~\ref{ladder}: for each meson, there are 3 contributing flavours ($u\overline{u}$, $d\overline{d}$ and $s\overline{s}$), enhancing the amplitude by a factor of 9, while the wavefunction normalisation in (\ref{eta0}) only gives a factor of 1/3. The amplitude squared is therefore enhanced by a factor of $3^2$ and we may therefore expect, in regions of phase space where the perturbative formalism is applicable, the $\eta\eta$ CEP cross section, despite the mixing angle suppression, to be dominant over $\pi\pi$ CEP -- we shall see in Section \ref{plots} that this is indeed the case. 

As the $\eta'$ state is dominantly flavour singlet, we also predict that $\eta'\eta'$ CEP should be strongly enhanced relative to $\pi\pi$ and $\eta\eta$ production. We note that the $gg\to\eta\eta'$ process, which vanishes for the diagrams given in Fig.~\ref{gg} due to the orthogonality of the $\eta$ and $\eta'$ flavour wavefunctions, can also occur via the diagrams of Fig.~\ref{ladder} (a), although the cross section is predicted to be suppressed by a factor of $\sin^2\theta_P$ relative to $\eta'\eta'$ production. 

It also immediately follows from helicity conservation along the quark lines that the production of transversely polarized vector mesons cannot proceed via these diagrams (in the limit that the quark $q_t=0$). Moreover, for longitudinally polarised vector mesons, we find
\begin{align}\label{long0}
T_{++,00}^{\rm lad}=T_{--,00}^{\rm lad}&=(2a-1)\,T_{++}^{\rm lad.}\;,\\ \label{long2}
T_{+-,00}^{\rm lad}=T_{-+,00}^{\rm lad}&=(1-2a)\,T_{+-}^{\rm lad.}\;,
\end{align}
which are antisymmetric under the interchange $x\leftrightarrow (1-x)$ (or $y\leftrightarrow (1-y)$), and will therefore vanish upon integration over the (symmetric) meson wavefunction, $\phi(x)$. This result for these `ladder' diagrams, where each meson state couples separately to two gluons, recalls the well-known Landau-Yang theorem~\cite{LY}, which states that a spin-1 particle cannot couple to two on-shell massless vector bosons\footnote{In the hard exclusive formalism, the $V\to gg$ amplitude is proportional to $(2x-1)$, and will therefore vanish upon integration over $\phi(x)$.}: although the intermediate $t$-channel gluon is in general far off-shell, the overall amplitude nevertheless vanishes. It therefore follows from (\ref{Tv1}) and  (\ref{long0})-(\ref{long2}) that the CEP of light vector pairs will be strongly suppressed in the perturbative regime, irrespective of their flavour structure, and indeed the $\rho\rho$, $\omega\omega$ and $\phi\phi$ rates are predicted within this formalism to be the same up to small (higher order, higher twist) corrections. 

Finally, we should in general consider the possibility of a two-gluon Fock component $|gg\rangle$ to the $\eta(\eta')$ mesons (see for example~\cite{Kroll02,Mathieu09}). The lowest order ($\alpha_S^0$) diagram, where the incoming gluons couple directly to the meson states, with an intermediate $t$-channel gluon exchanged between the mesons, will be strongly suppressed by the virtuality of the exchanged gluon, that is by the $\eta(\eta')\to gg^*$ form factor, where the $g^*$ is far off-mass-shell. In the hard exclusive formalism outlined in the previous sections, these form factors are modeled perturbatively: a representative diagram of the valence gluon contribution to the $gg \to \eta(\eta') \eta(\eta')$ process is shown in Fig.~\ref{ladder} (b). Thus it occurs to the same order in $\alpha_S$ as the valence quark process, with the size of the gluon contribution determined by the size of the meson two-gluon wavefunction $\phi_{g}(x)$. Unfortunately, the size of the two-gluon component of the $\eta(\eta')$ wavefunction is very poorly determined from experiment, and there exists no firm theoretical consensus as to its relative importance\footnote{We are very grateful to V. Chernyak and A. Groznin for a discussion of these issues.}. While as $Q^2\to \infty$, it can be shown that the $gg$ wavefunction vanishes due to QCD evolution~\cite{Ohrndorf81,Baier81}
\begin{equation}
\lim_{Q^2 \to \infty} \phi_g(x)=0\;,
\end{equation}
there is no reason to assume this will be the case at the experimentally relevant energies. In~\cite{Kroll02}, to consider one example, a fit to the $\eta(\eta')\gamma$ transition form factor, $F_{\eta(\eta')}(Q^2)$, suggests that the size of the $gg$ wavefunction {\it may} be of the same order as the quark contribution, although the extracted value is consistent with zero. Moreover, this fit assumes that the $q\overline{q}$ wavefunction is close to the asymptotic form, which appears to contradict the more recent BELLE data~\cite{Aubert09,Druzhinin09}, see also~\cite{Chernyak09}. There is also some indication from $\chi_c$ decays that the role/contribution of the `$gg$' component may be smaller than that found in~\cite{Kroll02}. In particular, no enhancement in the $\chi_{c0}\to \eta'\eta'$ relative to the $\chi_{c0}\to \eta\eta$ branching is observed, while the $\chi_{c2}\to \eta'\eta'$ decay is in fact strongly suppressed~\cite{Nakazawa04}: this effect would be quite surprising if a sizeable `$gg$' component to the $\eta'$ wavefunction were present.  While various other estimates are also available in the literature (see for example,~\cite{Thomas07,Ke11,Mathieu09,Ambrosino06,Escribano07}) no firm consensus exists about the precise size of the $gg$ contribution.

 Without a reliable input for $\phi_g(x)$, we therefore do not consider a full numerical calculation of the $gg$ contribution here, although this can readily be performed using the formalism outlined in the preceding sections. As described above, the relevant perturbative $gg\to 4g$ amplitude can be calculated in the usual way\footnote{The relevant $gg \to ggq\overline{q}$ amplitudes should in general be included as well.} and, as this does not vanish for $J_z=0$ initial-state gluons, we may expect it to enhance the $\eta\eta$ and $\eta'\eta'$ CEP rates (although, as with the $|q\overline{q}\rangle$ flavour singlet contribution, we will expect the $gg$ contribution to the $|\eta\rangle$ state to be fairly small). This will depend sensitively on the size of the two-gluon wavefunction: therefore, by considering the CEP of $\eta(\eta')$ pairs at sufficiently high invariant mass, it may be possible to extract some information about the relative importance of the leading-twist quark and gluon wavefunctions.

\section{$M\overline{M}$ CEP: results}\label{plots}

To calculate the $M\overline{M}$ production cross sections, we use the formalism described in Section~\ref{CEPform}. We take $f_\pi=133$ MeV and $f_\rho^\perp=f_\rho^0=200$ MeV, as in~\cite{Benayoun89}, and assume the flavour octet and singlet decays constants for states (\ref{eta0}) and (\ref{eta8}) are given by $f_0=f_8=f_\pi$: although this will in general not be true~\cite{Feldmann98}, we consider any deviation from this (which occurs at the $10-20\%$ level) to be within the uncertainties of the calculation. We assume in all cases a universal meson wavefunction given by (\ref{CZ}).

\begin{figure}[h]
\begin{center}
\includegraphics[scale=0.65]{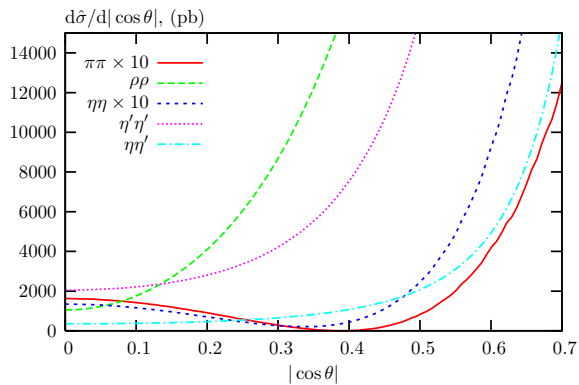}
\caption{The differential cross section ${\rm d}\hat{\sigma}/{\rm d}|\cos(\theta)|$ for the spin averaged $gg\to M\overline{M}$ process for various meson states.}\label{ang}
\end{center}
\end{figure}

\begin{figure}[h]
\begin{center}
\includegraphics[scale=0.65]{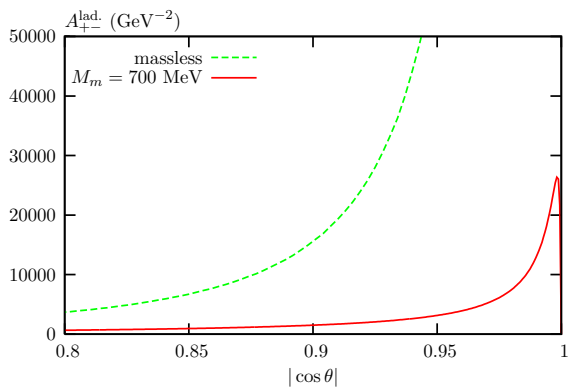}
\caption{$gg\to M\overline{M}$ amplitude for production of flavour singlet mesons (\ref{eta0}), with gluons in a $|J_z|=2$ state. Plotted is the amplitude as a function of $|\cos(\theta)|$ for massless quarks, and with the inclusion of a non-zero quark mass. For the purpose of illustration we choose the value of $m_q=0.35$ GeV, but it should be noted that the amplitude vanishes at $|\cos \theta|=1$ for any non-zero quark mass.}\label{mquark}
\end{center}
\end{figure}

We first show in Fig.~\ref{ang} the spin-averaged differential cross section ${\rm d}\hat{\sigma}/{\rm d}|\cos(\theta)|$ for the $gg\to M\overline{M}$ subprocess (the distributions are of course symmetric under the interchange $\cos \theta \to -\cos \theta$). In all cases the cross sections are strongly enhanced as $|\cos \theta|\to 1$, where they are in fact divergent in the limit of massless quarks taken here. Clearly, in the limit of small angle scattering we can no longer trust the calculation, but out to reasonable values of $\cos\theta$ we expect the dominant angular behaviour to be described by these distributions: indeed measurements at BELLE of $\gamma\gamma \to \pi^+\pi^-$ (and $\gamma\gamma \to K^+K^-$, which is expected theoretically to have a very similar angular distribution) out to $|\cos\theta|=0.6$~\cite{Nakazawa04} and $\gamma\gamma \to \pi^0\pi^0$ out to $|\cos\theta|=0.8$~\cite{:2009cka} observe the $1/\sin^4\theta$ distribution predicted within the perturbative framework. It should also be noted that, upon the inclusion of non-zero quark masses, the amplitudes are not in fact divergent as $|\cos \theta|\to 1$. An interesting example of this is in the case of the $|J_z|=2$ flavour singlet amplitude (\ref{lad2}), which, due to conservation of the projection of the total angular momentum, $J$, onto the $z$-axis, must vanish in the limit of zero angle scattering. While for massless quarks this is lost due to the IR divergence of the quark propagators, upon the inclusion of a non-zero quark mass the amplitude does indeed vanish at $|\cos\theta|=1$, see Fig.~\ref{mquark}. Although we do not show it explicitly here, a similar result must of course hold for the nonsinglet amplitude (\ref{T+-}), and we note that the non-zero $gg\to V\overline{V}$ helicity amplitudes (\ref{Tv2}, \ref{Tv3}) vanish as expected in the forward or backward limit, when the initial and final-state spin projections on the $z$-axis are opposite. Nonetheless, we of course cannot trust fixed-order perturbation theory in this region of phase space.

\begin{figure}[h]
\begin{center}
\includegraphics[scale=0.6]{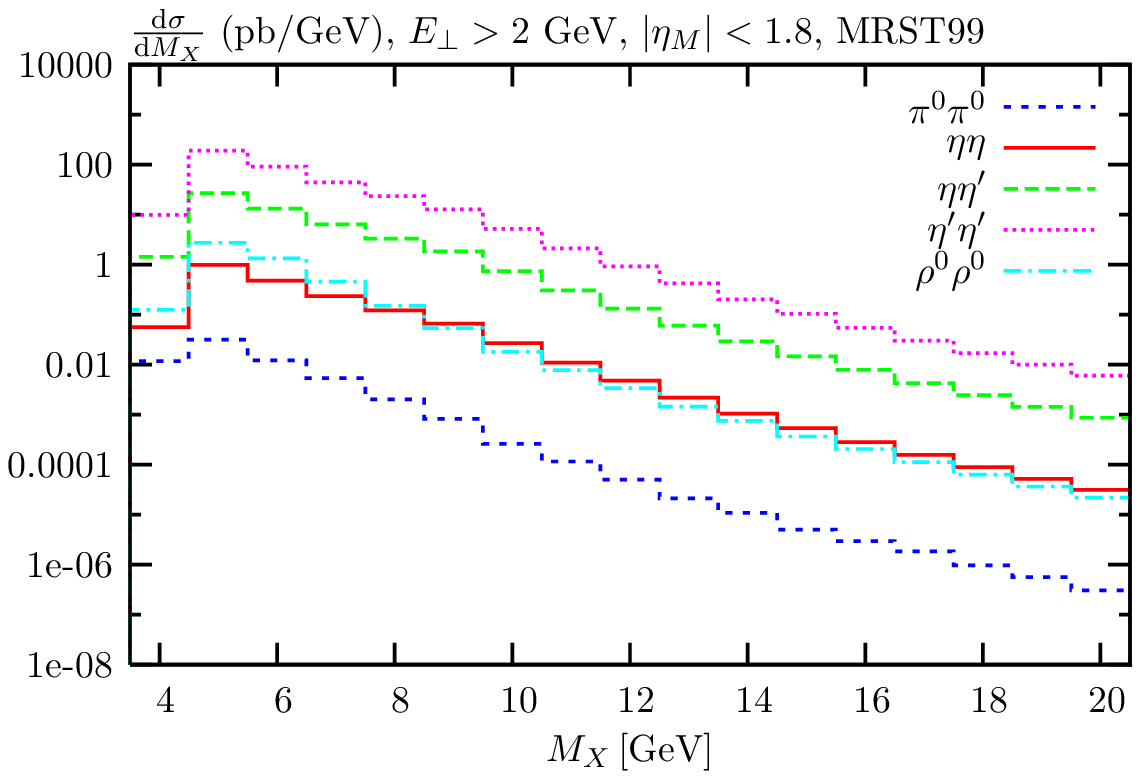}
\includegraphics[scale=0.6]{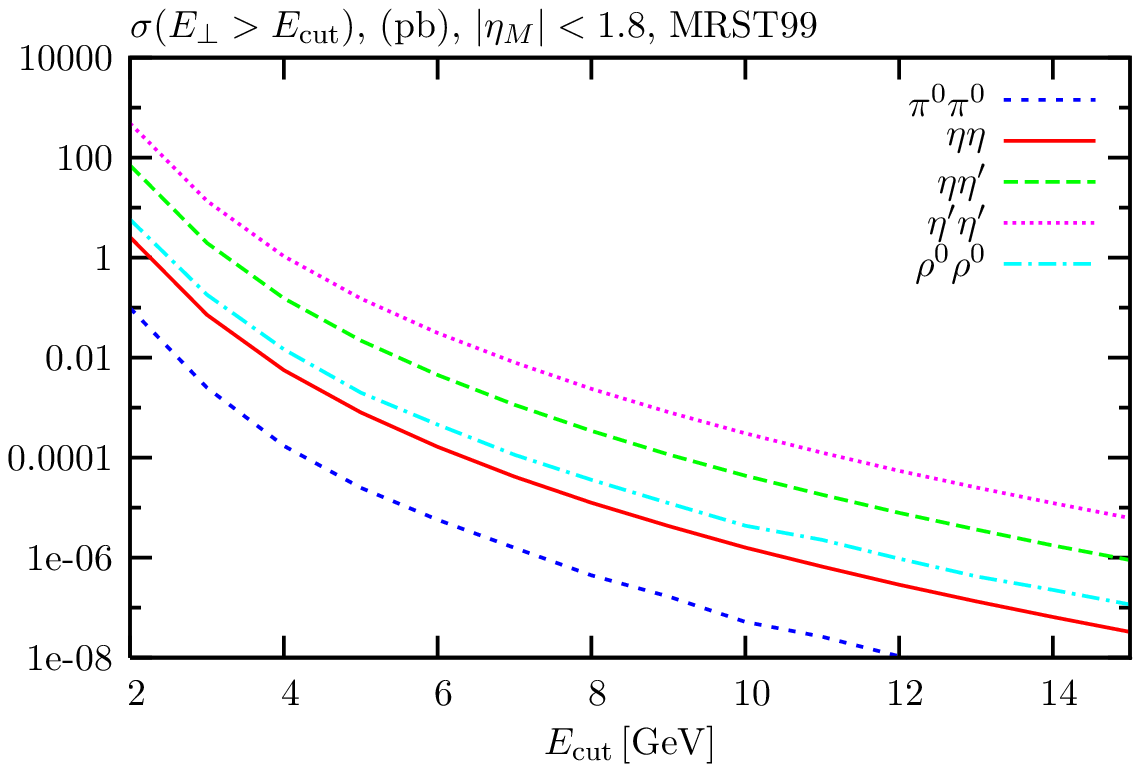}
\caption{${\rm d}\sigma/{\rm d}\ln M_X^2$ for meson transverse energy $E_\perp>2$ GeV, and cross section as a function of the cut $E_{\rm cut}$ on the meson $E_\perp$ at $\sqrt{s}=1.96$ TeV for the CEP of meson pairs, calculated within the perturbative framework, including both the $J_z=0$ and $|J_z|=2$ contributions. The mesons are restricted to have pseudorapidity $|\eta_M|<1.8$. Meson masses are neglected throughout.}\label{vsm}
\end{center}
\end{figure}

It should be noted however that the angular distributions in Fig.~\ref{ang}, which are averaged over all incoming gluon helicities at the cross section level, are not directly relevant to the case of meson pair CEP, and are only shown for illustration. In particular, we have noted the importance of the $J_z=0$ selection rule that operates for CEP. This has significant consequences for the case of $M\overline{M}$ CEP: in particular, the subprocess amplitudes for the flavour-nonsinglet scalar (\ref{T+-}) and all vector (\ref{Tv1}) states vanish for $J_z=0$ gluons, and we therefore expect the CEP cross sections to be heavily suppressed, see (\ref{simjz2}), relative to the cross section for flavour-singlet states ($\eta'\eta'$, $\eta\eta$, $\eta\eta'$), where the subprocess amplitudes do not vanish for $J_z=0$ incoming gluons, see (\ref{lad0}). The $\eta\eta$ and $\eta\eta'$ CEP cross sections are strongly dependent on the precise level of $\eta-\eta'$ mixing, through which a $J_z=0$ component can enter. In Fig.~\ref{vsm} we show the CEP cross sections ${\rm d}\sigma/{\rm d}M_X$, where $M_X$ is the invariant mass of the meson pair, for the production of various scalar and vector states. The suppression of the $\pi^0\pi^0$ and vector meson cross sections is clear\footnote{The charged $\pi^+\pi^-$ and $\rho^+\rho^-$ CEP cross sections are expected to be a factor of 2 larger from isospin symmetry and the non-identity of the final state particles.}, in particular in the $\pi\pi$ case where the radiative zero in the angular distribution will tend to further reduce the cross section. The $\eta'\eta'$ cross section, on the other hand, is predicted to be quite large. The vector meson $\rho\rho$ cross section is also shown: within the perturbative formalism, the $\phi\phi$ and $\omega\omega$ rate are to lower order expected to be identical, see Section~\ref{svect}. 

\begin{figure}[h]
\begin{center}
\includegraphics[scale=0.6]{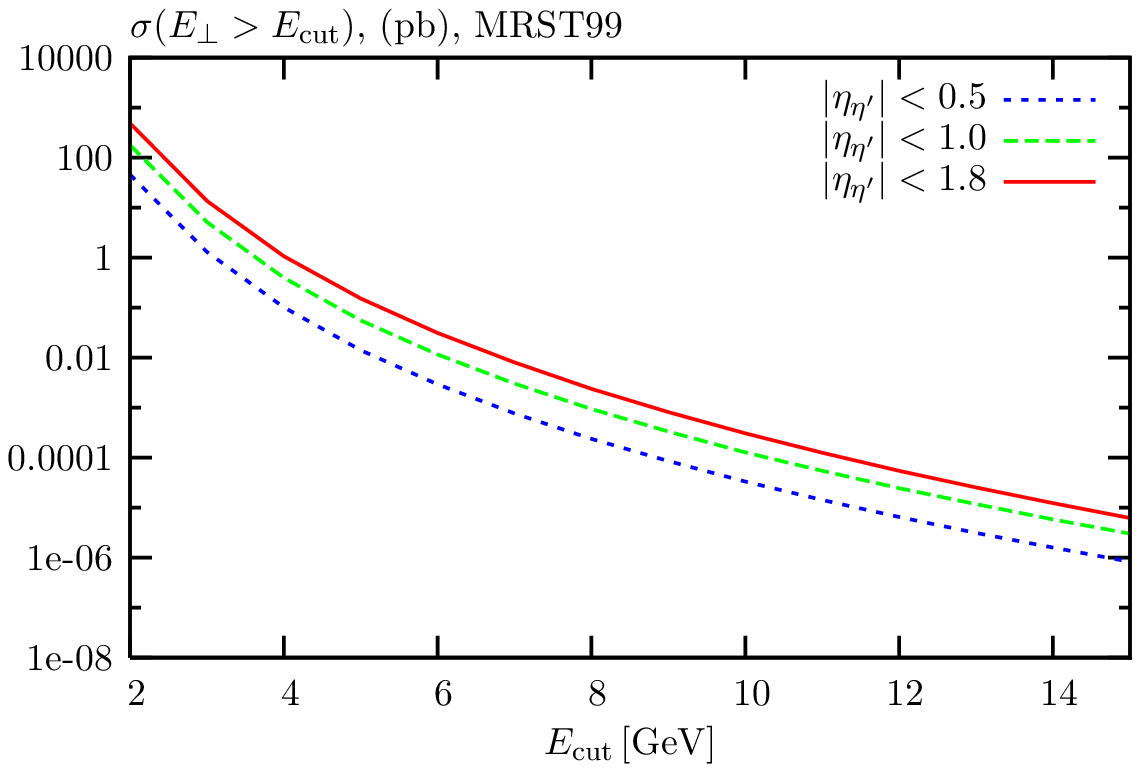}
\includegraphics[scale=0.6]{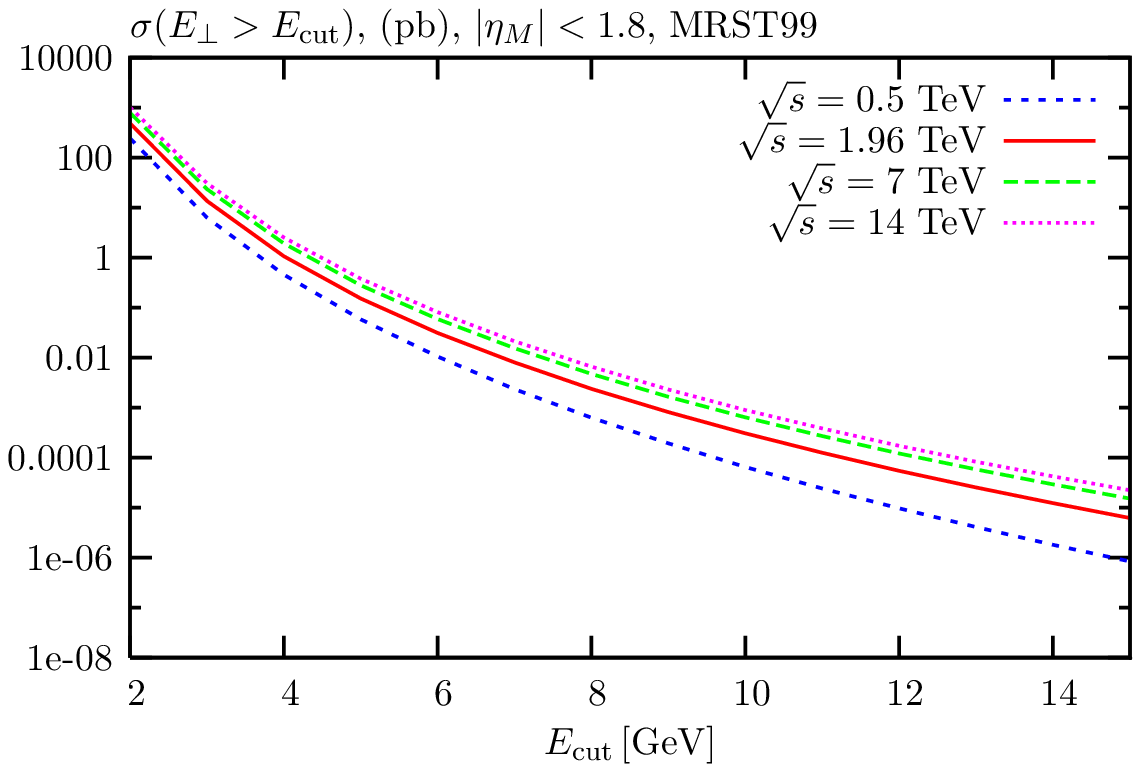}
\caption{$\eta'\eta'$ CEP cross section at $\sqrt{s}=1.96$ TeV, as a function of the cut $E_{\rm cut}$ on the meson $E_\perp$, calculated within the perturbative framework, including both the $J_z=0$ and $|J_z|=2$ contributions. Plotted is the cross section for different values of c.m.s energy $\sqrt{s}$ and cut on the meson pseudorapidity $\eta_M$. Meson masses are neglected throughout.}\label{vset1}
\end{center}
\end{figure}

\begin{figure}[h]
\begin{center}
\includegraphics[scale=0.6]{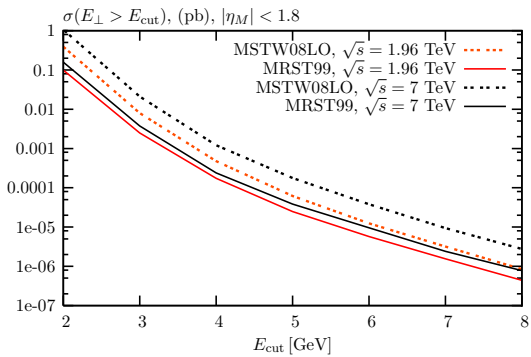}
\caption{$\pi^0\pi^0$ pair CEP cross section as a function of the cut $E_{\rm cut}$ on the meson $E_\perp$, at $\sqrt{s}=1.96$ and $\sqrt{s}=7$ TeV, calculated within the perturbative framework, including both the $J_z=0$ and $|J_z|=2$ contributions, for two choices of PDF, MSTW08LO and MRST99.}\label{pdfc}
\end{center}
\end{figure}

In Fig.~\ref{vsm} we also show the CEP cross sections for the production of the same states, as a function of the cut, $E_{\rm cut}$, on the meson transverse energy $E_\perp =E \sin\theta$ (where $\theta$ is the angle of the pion from the beam axis in the lab frame), and the same effects are clear. In Fig.~\ref{vset1}, we show the effect of increasing the c.m.s. energy, $\sqrt{s}$, and cut on the meson pseudorapidity, $\eta_M$, on the cross section, taking the case of $\eta'\eta'$ CEP for illustration. Unfortunately, for the low $x$ and $Q^2$ values probed in the CEP of lighter mass objects, there is a large degree of uncertainty in the single PDFs. Recalling the strong PDF dependence ($\sigma_{\rm CEP} \sim (xg)^4$) of the CEP cross section, the effect of this will be quite severe. This is already relevant at Tevatron energies, but is a stronger source of uncertainty when it comes to making predictions for the LHC, where the probed $x$ values are even lower. In Fig.~\ref{pdfc} we show this explicitly by comparing the predicted $\pi^0\pi^0$ CEP cross section as a function of $E_{\rm cut}$ for two different choices of PDF set, MSTW08LO~\cite{Martin:2009iq} and MRST99~\cite{Martin:1999ww}, as in~\cite{HarlandLang10} for the case of $\gamma\gamma$ CEP. Clearly this is a significant source of uncertainty in the normalisation of the predicted cross sections, with the MSTW08LO and MRST99NLO sets providing approximate upper and lower bounds, respectively, on the range of predictions coming from different PDF sets. We use these two particular sets because we would argue that they span a realistic range of small $x$ parton distributions. MSTW08LO is the outcome of a leading-order pQCD global fit to an up-to-date set of DIS and other hard scattering data. At small $x$, it gives a reasonable, though not perfect, description of HERA $F_2(x,Q^2)$ data. The corresponding NLO version (MSTW08NLO) has a very different gluon distribution at small $x$ and $Q^2$, with $g(x,Q_0^2) < 0$ for $ x \lesim 10^{-2}$. Because of this behaviour, the MSTW08NLO gluon gives unstable results when used to calculate the skewed PDF, $f_g$. We prefer to use the older MRST99 NLO set, which has a more benign small-$x$ form, while still retaining the essential features of a NLO fit, in particular a gluon that is smaller at small $x$ than at LO. We take the MRST99 set, which gives predictions that are in better agreement with the existing Tevatron data, as our default set, see~\cite{HarlandLang10} for more discussion of these issues.

As was discussed in the Introduction, the  calculation of the $\pi^0\pi^0$ CEP cross section has important consequences for the possible $\pi^0\pi^0$ background
to $\gamma\gamma$ CEP, when one photon from each $\pi^0$ decay is undetected or the two photons merge. At first sight it would appear that the cross section for this purely QCD process could be much larger than that in the $\gamma\gamma$ channel and so would constitute an appreciable background, but fortunately this is not the case. Firstly, we have seen that the amplitude to form an exclusive pion with large transverse momentum, $k_\perp$, is proportional to the ratio $f_\pi/\sqrt{\hat{s}}\sim f_\pi/k_\perp$ (see (\ref{wnorm}) and (\ref{T+-})), that is the cross section of the $gg\to \pi^0\pi^0$ hard subprocess contains the numerically small factor $(f_\pi/k_\perp)^4$ which in the region of interest is comparable (or even smaller) to the QED suppression, $(\alpha_{QED}/\alpha_S)^2$, of the $gg\to\gamma\gamma$ cross section. Secondly, and crucially, the vanishing of the LO amplitude $gg\to\pi^0\pi^0$ with $J_z=0$ initial-state gluons leads to a further $\sim$ two orders of magnitude suppression in the CEP cross section. We therefore expect the $\pi^0\pi^0$ background contribution to $\gamma\gamma$ CEP to be small\footnote{As noted in Section~\ref{bg}, the cross section may however be somewhat enhanced by higher-twist or NNLO corrections which allow a $J_z=0$ component. At lower values of $\pi\pi$ invariant mass, an additional `non-perturbative' contribution must also be considered, see Section~\ref{nsect}.} -- we shall address this issue in more detail in a forthcoming publication~\cite{HarlandLangfut}.

Our results can readily be extended to the kaon sector, although we do not consider this numerically here. In particular, under the assumption of exact $SU(3)$ flavour symmetry, the $K^0\overline{K}^0$ and $K^+K^-$ CEP cross sections can be calculated in the same way as the $\pi^0\pi^0$ and $\pi^+\pi^-$ cross sections, with the replacement $f_\pi \to f_K$. In fact, $SU(3)$ flavour symmetry breaking effects can be non-negligible, and to precisely estimate the $K^0\overline{K}^0$ and $K^+K^-$ cross sections, a modified narrower form of the meson wavefunction, which accounts for asymmetry between the $s$ and $(u,d)$ quark masses, should be taken, see~\cite{Benayoun89} for more details. Without the inclusion of this modified wavefunction, the perturbative formalism tends to overestimate the $\gamma\gamma \to K^+ K^-$ cross section when compared to BELLE data~\cite{Nakazawa04}.

Finally, we note that the CEP of these states is now included in the publicly available SuperCHIC Monte Carlo~\cite{SuperCHIC}, which also generates the CEP of a range of low-mass states ($\chi_{c,b}$, $\eta_{c.b}$, $\gamma\gamma$).

\section{Non-perturbative meson pair production}\label{nsect}

\begin{figure}[h]
\begin{center}
\includegraphics[scale=0.9]{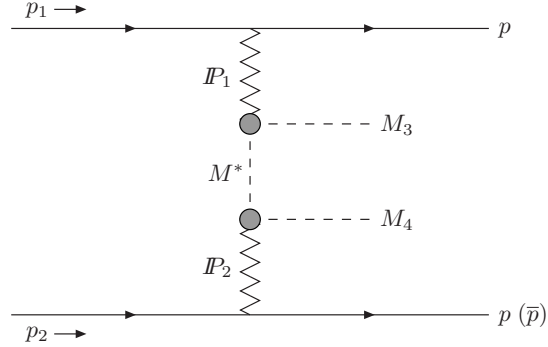}
\caption{Non-perturbative meson pair ($M_3$, $M_4$) CEP mechanism, where $M^*$ is an intermediate off-shell meson of type $M$.}\label{npip}
\end{center}
\end{figure}

For low values of the meson pair invariant mass we may not expect the perturbative framework described in Sections~\ref{back} and \ref{bg} to be applicable. In this region of phase space we can model the process using a `non-perturbative' picture~\cite{Pumplin76,Azimov74,Lebiedowicz09} shown in Fig.~\ref{npip}: the $M\overline{M}$ pair are created via double-Pomeron exchange, with an intermediate $t$-channel off-shell meson, $M^*$.

We can calculate the resultant amplitude using the standard tools of Regge theory, see for example~\cite{regge}. The CEP cross section is given by
\begin{equation}\label{ncross}
\sigma^{CEP}=\frac {S^2}{16\pi(16\pi^2)^2}\int dp^2_{1\perp}dp^2_{2\perp}dy_1dy_2dk^2_{\perp}\frac{|\mathcal{M}|^2}{s^2}\;,
\end{equation}
where $\sqrt{s}$ is the c.m.s. energy, $p_{1\perp}, p_{2\perp}$ are transverse momenta of the outgoing protons and $k_\perp$ is the meson transverse momentum. $S^2$ is the soft survival factor, expected to be of a similar size to $S^2_{\rm eik}$ in the perturbative case (\ref{ampnew}). The matrix element is given by $\mathcal{M}=\mathcal{M}_{\hat{t}}+\mathcal{M}_{\hat{u}}$, with $\hat{t}=(P_1-k_3)^2$, $\hat{u}=(P_1-k_4)^2$, where $P_i$ is the momentum transfer through Pomeron $i$, and $k_{3,4}$ are the meson momenta. We have
\begin{equation}\label{namp}
\mathcal{M}_{\hat{t}}=\frac 1{M^2-\hat{t}} F_p(p^2_{1\perp})F_p(p^2_{2\perp})F^2_M\sigma_0^2
\bigg(\frac{s_{13}}{s_0}\bigg)^{\alpha(p_{1\perp}^2)}\bigg(\frac{s_{24}}{s_0}\bigg)^{\alpha(p_{2\perp}^2)}\;,
\end{equation}
where $M$ is the meson mass and we take $s_0=1\,{\rm GeV}^2$ and $\alpha_P(p^2_{i\perp})=1.08-0.25|p^2_{i\perp}|$, for $p^2_{i\perp}$ measured in ${\rm GeV}^2$~\cite{DL92}, and $s_{ij}=(p_i'+k_j)^2$ is the c.m.s. energy squared of the final state proton-meson system $(ij)$. The proton form factors are as usual taken to have an exponential form, $F_p(t_i)=\exp(B_it_i/2)$, while the Pomeron trajectory is included in the definition of the slope
\begin{equation}
B_i=b_0+2\alpha'\log\bigg(\frac{s_{ij}}{s_0}\bigg)\;,
\end{equation}
with $b_0=4\,\,{\rm GeV}^{-2}$. Concentrating on the case of $\pi^0\pi^0$ production, the overall cross section normalisation is set by the total pion--proton cross section $\sigma(p\pi)=\sigma_0 (s_{ij}/s_0)^{\alpha(0)-1}\approx30$ mb at the relevant sub-energy. 

Finally, we have to include an additional suppression factor to calculate the genuinely exclusive $M\overline{M}$ cross section -- the small probability not to produce any additional hadrons in the Pomeron-Pomeron fusion process, which can be evaluated by taking a Poisson distribution in the multiplicity of secondaries. Note however that we cannot assume a Poisson distribution in the multiplicity of all charged tracks, as this ignores the correlation $N^+=N^-$. Secondly, a fraction of the secondary pions is produced via resonance decay or even in a minijet fragmentation -- in the latter case we should consider a Poisson distribution in the number of minijets. Finally, besides charged particles there are also neutral secondaries to consider. For the above reasons we {\it assume} a Poisson distribution in the number of negatively charged  particles. The mean multiplicity, $n=\langle N^-\rangle$, grows with the Pomeron-Pomeron energy, $\hat{s}=(P_1+P_2)^2$, as $n\simeq c\cdot\ln(\hat s/s_0)$,  with the  coefficient\footnote{It would be interesting to measure the height of the `plateau', $dN/d\eta$, produced in Pomeron-Pomeron (DPE) collisions at the LHC.} $c\sim 0.5 -1$. The Poissonian probability not to create any additional secondaries is then given by $w=\exp(-n)= (\hat{s}/s_0)^{-c}$. For numerical estimates we take $c=0.7$ here.

One may ask what the origin of this additional suppression, $w$, is in terms of the non-perturbative diagrams of the type shown in Fig.~\ref{npip}.  We should recall that this is just a representative diagram: besides `eikonal' proton-proton rescattering (not shown in Fig.~\ref{npip}, but included as the soft gap survival factor
$S^2=S^2_{\rm eik}$ in (\ref{ncross})) we have to account for the possibility of additional meson-proton (and meson-meson) interactions which play the role of the survival
factor $S^2_{\rm enh}$ shown in Fig.~\ref{fig:pCp}, and for the reggeization of the virtual $t$-channel meson $M^*$. It is this reggeization of the meson $M^*$ trajectory which leads to the additional power suppression described above. In the case of $\pi\pi$ CEP for example, with the slope of the pion trajectory taken as $\alpha'_\pi=0.9$ GeV$^{-2}$, the coefficient $c=0.7$ corresponds to a mean momentum transferred $-t=0.4$ GeV$^2$. Actually, here we are considering the case of pions with a rather large $k_\perp> 1\ -\ 2$ GeV, for which the contribution to the diagram shown in Fig.~\ref{npip} from pion reggeization, taken literally, would be negligible. However besides the pion contribution in the place of $M^*$ there may be $a_2$ and/or $a_1$ Regge trajectories, and the whole structure of this non-perturbative interaction is therefore very complicated. To get an order of magnitude estimate of this effect we consider the simplified diagram of Fig.~\ref{npip}, neglecting the pion reggeization but including the probability, $w$, not to produce any additional secondaries, the procedure for calculating which we have described above.

We note that the CEP of $\pi\pi$ pairs within the non-perturbative framework was considered recently in~\cite{Lebiedowicz11} (see also~\cite{Lebiedowicz09,Staszewski:2011bg}). The model used is largely similar to that considered here, but this extra suppression is not included and so their estimates for the $\pi\pi$ non-perturbative cross section in general become significantly larger than our predictions as $M_{\pi\pi}$ is increased, leading for example to roughly an order of magnitude difference at the
$M_{\pi\pi}\sim M_\chi$ mass.\footnote{In the previously measured mass range ($M_{\pi\pi}\lesssim 2$ GeV) at the CERN ISR~\cite{Breakstone90}, where the effect of this suppression is not too large, our results are roughly consistent, when we account for the secondary Reggeon contributions included in~\cite{Lebiedowicz09}, but which we have omitted above for simplicity, as their contribution will be insignificant at the c.m.s. values we consider here.} Moreover, the authors of~\cite{Lebiedowicz11} do not consider the possible perturbative contribution due to the $gg\to \pi\pi$ subprocess, which we expect to be dominant, and therefore increase the overall rate, for higher values of $M_{\pi\pi}$. In fact, these two differences may compensate each other in the $M_{\pi\pi}\sim M_\chi$ mass region, leading to an approximate consistency between our results.

\begin{figure}
\begin{center}
\includegraphics[scale=0.6]{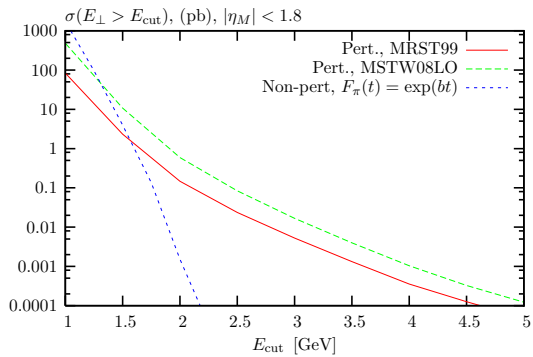}
\caption{Comparison between perturbative (\ref{bt}) and non-perturbative (\ref{namp}) $\pi^+\pi^-$ CEP cross sections as a function of the cut $E_{\rm cut}$ on the pion transverse energy $E_\perp$. In the non-perturbative case, an exponential pion form factor $F_\pi(t)=\exp(bt)$, with $b=1\,{\rm GeV}^{-2}$ is taken.}\label{npi}
\end{center}
\end{figure}

\begin{figure}
\begin{center}
\includegraphics[scale=0.6]{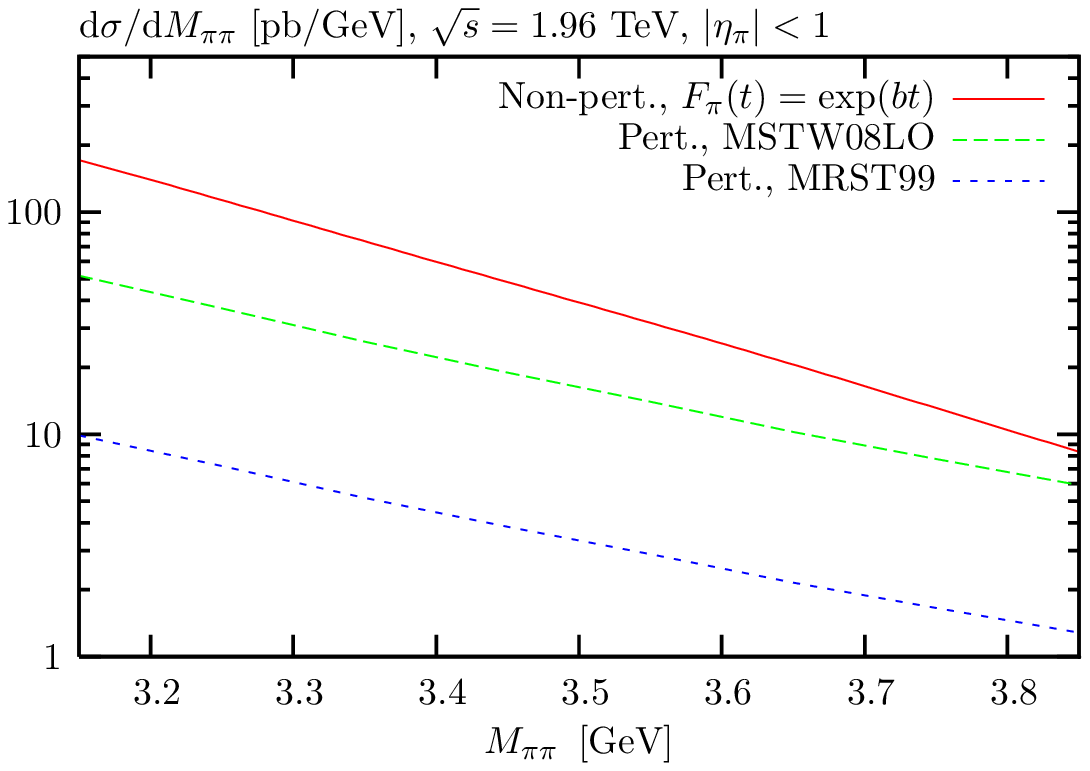}\qquad
\includegraphics[scale=0.6]{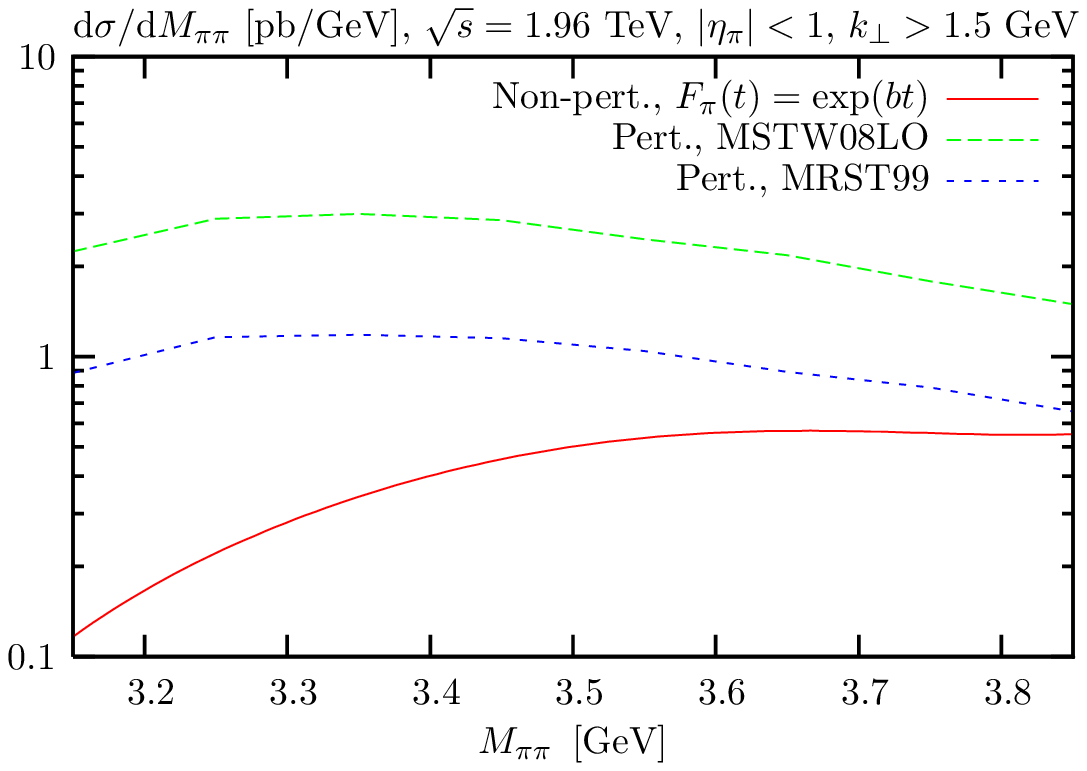}
\caption{Comparison between perturbative (\ref{bt}) and non-perturbative (\ref{namp}) $\pi^+\pi^-$ CEP differential cross sections ${\rm d}\sigma/{\rm d}M$, for different cuts on the pion transverse momentum, $k_\perp$, and for the pion pseudorapidity restricted to lie in the region $\eta<1$. In the non-perturbative case, an exponential pion form factor $F_\pi(t)=\exp(bt)$, with $b=1\,{\rm GeV}^{-2}$ is taken.}\label{npim}
\end{center}
\end{figure}

In the case of non-perturbative $M\overline{M}$ production, large meson $k_\perp$ values are suppressed by the form factor, $F_M(\hat{t})$, of the intermediate off-shell meson in (\ref{namp}), although in this case it is not completely clear what form to take for $F_M(\hat{t})$. In particular, we may either take the `soft' exponential $\exp(b\hat{t})$, or the `hard' power-like form $\sim 1/\hat{t}$, both of which are used in the literature~\cite{Pumplin76,Desai78,Lebiedowicz09}. As $|\hat{t}|$ (that is, the meson $k_\perp^2$) is increased, these two choices of form factor give vastly different cross sections, with the power-like form factor giving a significantly larger rate. However, in general we should expect a smooth matching at a reasonably low value of $k_\perp$ between the non-perturbative and the perturbative region of phase space, where we expect the mechanism outlined in the previous sections to be dominant. At higher values of $k_\perp$, the contribution from the power-like form factor should not be included, since it represents a form of double-counting of the perturbative tail region, which we should calculate within the pQCD CEP formalism. We show this in Fig.~\ref{npi}, where the non-perturbative (with exponential form factor) and perturbative $\pi^+\pi^-$ CEP cross sections are plotted as a function of the cut on the $E_\perp$ of the pions: the matching between the two regions occurs at the reasonable value of $E_\perp \sim 1.5$ GeV. Due to the strong $J_z=0$ suppression of the perturbative cross section, the transition region occurs at a somewhat higher value of pion $E_\perp$ (and hence $M_{\pi\pi}$) than we would otherwise find, and this has the consequence that both the perturbative and non-perturbative mechanisms may be relevant in the $M_{\pi\pi}\sim M_\chi$ mass region relevant for evaluating the potential continuum background to $\chi \to \pi\pi$ CEP. Plotted in Fig.~\ref{npim} is the differential cross section ${\rm d}\sigma/{\rm d}M_{\pi\pi}$ for the perturbative and non-perturbative contributions in the mass region relevant to $\chi_c$ CEP at $\sqrt{s}=1.96$ and $|\eta_\pi|<1$. When no cut is placed on the pion transverse momentum the non-perturbative mechanism is predicted to be dominant in the $\chi_c$ mass region, however when more realistic experimental cuts are placed on the pion $k_\perp$, the relative contribution of the non-perturbative mechanism is suppressed by the pion form factor, $F_\pi(t)$: this can be seen for example in the case of the cut $k_\perp>1.5$ GeV, for which the perturbative mechanism is predicted to be dominant. We note that, when available, CDF data~\cite{malbrow,Albrow:2010zz} on central exclusive $\pi^+\pi^-$ production may provide useful information about the off-shell pion form factor and help further illuminate this issue. 

In fact, the CEP mechanism shown in Fig.~\ref{fig:pCp} does not represent the true perturbative tail to the double Pomeron exchange process shown in the Fig.~\ref{npip}: we shall consider this process in the following section. 

\section{$M\overline{M}$ CEP: secondary mechanism}\label{sfors}

\begin{figure}[h]
\begin{center}
\subfigure[]{\includegraphics[clip,trim=0 15 0 0,scale=0.5]{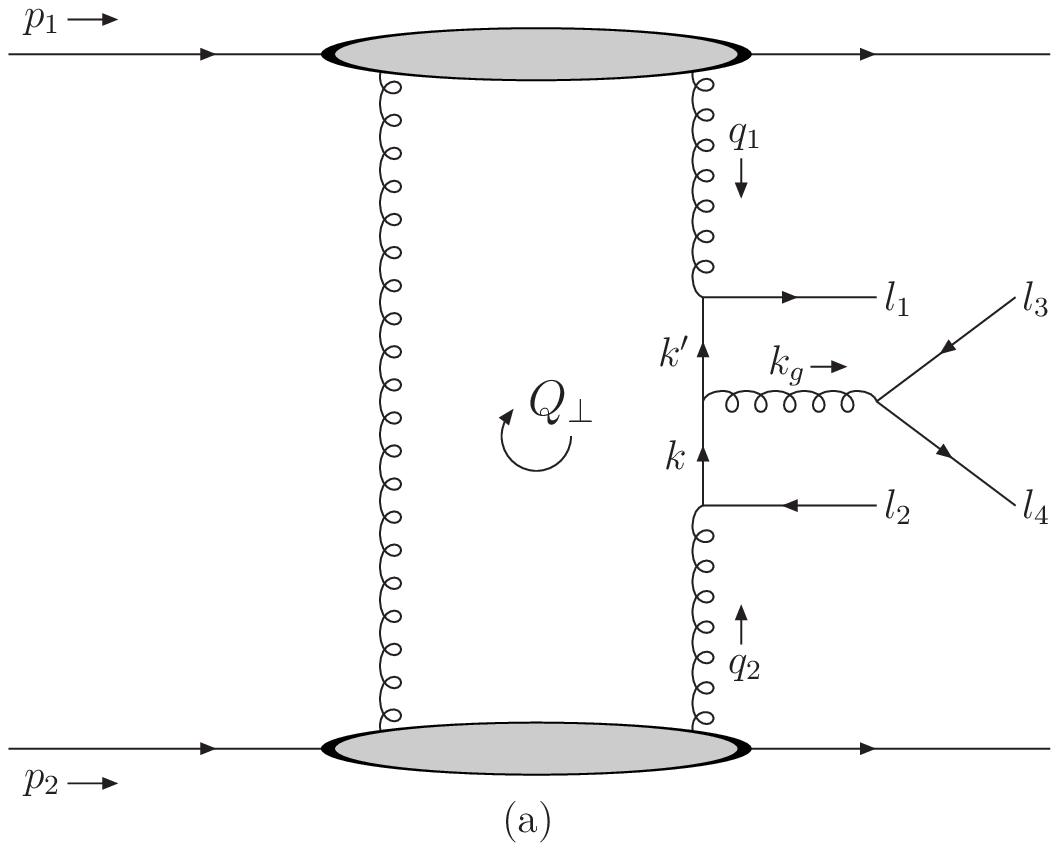}}\qquad
\subfigure[]{\includegraphics[clip,trim=0 15 0 0,scale=0.5]{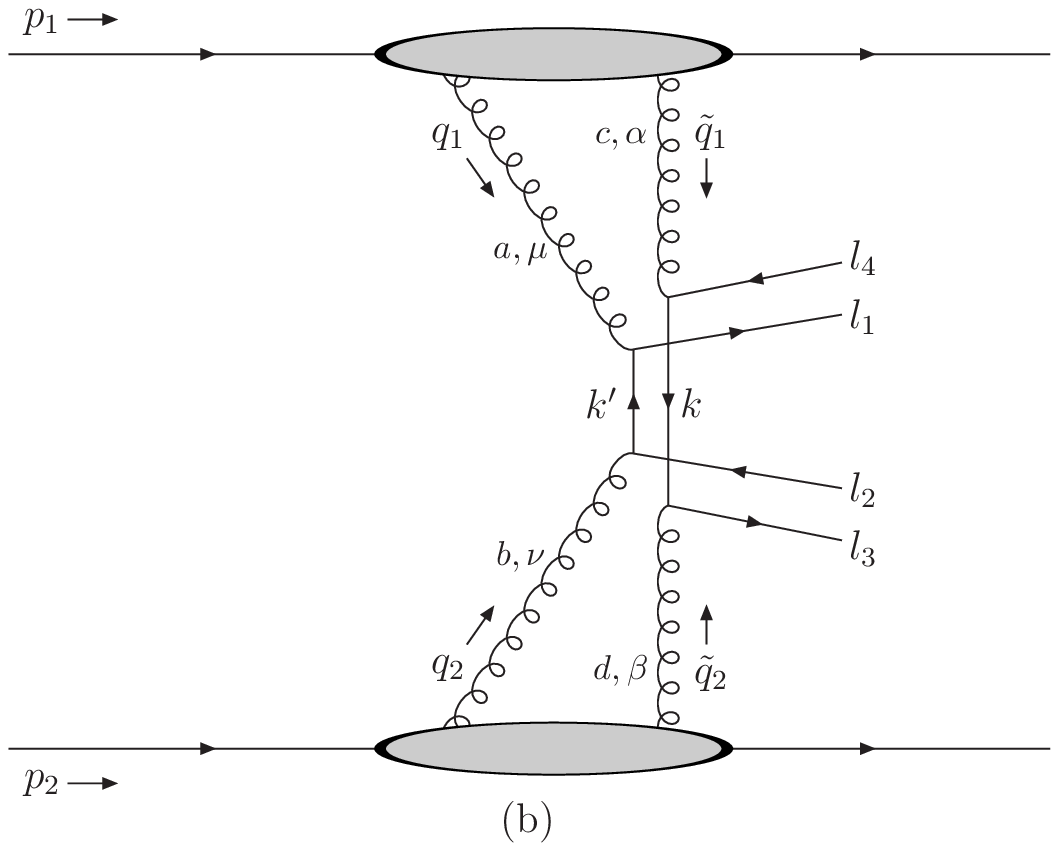}}
\caption{{\bf (a)} Representative perturbative diagram for `skewed' meson pair CEP. {\bf (b)} Representative perturbative diagram for `symmetric' meson pair CEP.}\label{fors}
\end{center}
\end{figure}

As well as the standard CEP diagram shown in Fig.~\ref{fors} (a), there is in general a second mechanism\footnote{The authors are grateful to Jeff Forshaw for bringing this type of diagram to our attention.} for producing meson pairs at high $k_\perp$. This is shown in Fig~\ref{fors} (b): the second $t$-channel gluon now couples directly to a quark line, with the collinear $q\overline{q}$ pairs forming mesons in the usual way, in both the flavour-nonsinglet and flavour-singlet combinations described in Sections~\ref{gg} and \ref{eta}, respectively. We shall label Figs.~\ref{fors} (a) and (b) as `skewed' and `symmetric' CEP, respectively: in Fig.~\ref{fors} (a) the second $t$-channel gluon is much softer than the fusing gluons, whereas in Fig.~\ref{fors} (b) both $t$-channel gluons participate symmetrically in the hard subprocess. The symmetric diagram represents the perturbative tail to the non-perturbative double Pomeron exchange mechanism shown in Fig.~\ref{npip}, resolving the two-gluon structure of the exchanged Pomerons. As this diagram is of the same order in $\alpha_S$, it might in principle give an important contribution to the meson pair CEP cross sections given in Section \ref{plots}. Moreover, in the case of scalar non-singlet (and vector meson) skewed CEP amplitude, which are suppressed by the $J_z=0$ selection rule, the effect may be particularly important: in Fig.~\ref{fors} (b), the $\pi\pi$ and vector meson production amplitudes will no longer vanish for forward outgoing protons.

The symmetric CEP amplitude, considering for simplicity the case of forward outgoing protons ($p_{i_\perp}=0$), can be written in the form

\begin{equation}\label{bsym}
T_{\rm sym.}=\pi^2 \int \frac{d^2 q_{1\perp}}{ q_{1\perp}^4 q_{2\perp}^4 }\,\mathcal{M}_{\rm sym.}\,f_g(x_1,\tilde{x}_1, q_{1_\perp}^2,\mu^2;t_1)f_g(x_2,\tilde{x}_2,q_{2_\perp}^2,\mu^2;t_2) \; ,
\end{equation}
where the notation follows from Fig.~\ref{fors} (b) and the subprocess amplitude $\mathcal{M}_{\rm sym.}$ is given by
\begin{align}\label{msym}
\mathcal{M}_{\rm sym.}&=\frac{4}{M_X^4}\frac{1}{N_C^2-1}\delta^{ac}\delta^{bd}q_{1\perp}^\mu q_{2\perp}^\nu q_{1\perp}^\alpha q_{2\perp}^\beta V^{abcd}_{\mu\nu\alpha\beta}\\
&\equiv \frac{4}{M_X^4} V_{\rm sym.}\;.
\end{align}
For example, considering the specific case of scalar non-singlet meson production, $V_{\rm sym.}$, as shown in Fig~\ref{fors} (b), is given by
\begin{equation}\label{spec}
V_{\rm sym.} = 16\pi^2 \alpha_S^2\frac{C_F}{2N_C}\int\, {\rm d}x\, {\rm d}y \,\phi(x)\phi(y)\,\frac{{\rm Tr}({\not}k{\not}q_{2\perp}{\not}k_4
{\not}q_{2\perp}{\not}k'{\not}q_{1\perp}{\not}k_3{\not}q_{1\perp})}{2k^2k'^2}\;.
\end{equation}
%
The other diagrams corresponding to interchanging the outgoing quark legs $l_1 \leftrightarrow l_2$ and $l_3 \leftrightarrow l_4$, as well as the $s$-channel diagrams containing 3-gluon vertices, should also be included, and the resulting expression can then be combined with (\ref{bsym}) to give an explicit evaluation of the full symmetric CEP amplitude. 

However, simply by inspecting the form of the amplitude (\ref{msym})-(\ref{spec}) we find that the symmetric process in general represents a power correction to the skewed CEP process, and therefore formally gives a subleading contribution. We demonstrate this in full in Appendix~\ref{dps}, where we show that
\begin{equation}\label{sup}
\frac{A_{\rm sym.}}{A_{\rm skew.}} \sim \frac{\langle Q_\perp^2 \rangle}{k_\perp^2}\;,
\end{equation}
where $k_\perp$ is the meson transverse momentum, and $\langle Q_\perp^2 \rangle$ is the average gluon transverse momentum in Fig.~\ref{fors}(b), see (\ref{qav}). That is, the symmetric amplitude is suppressed by the meson $k_\perp^2 \sim \hat{s}$, and its relative contribution will decrease as the meson $k_\perp$ is increased. The skewed CEP amplitude is therefore, at least formally, dominant at high $k_\perp$. However, if we take $k_\perp\approx M_X/2=5/2$ GeV we find
\begin{equation}\label{comp}
\frac{\sigma_{\rm sym.}}{\sigma_{\rm skew.}}\approx \frac{1}{4} - \frac{1}{2}\;,
\end{equation}
for a typical $\langle Q_\perp^2 \rangle= 3-4$ ${\rm GeV}^2$. At smaller values of $k_\perp$, it therefore appears that the suppression of the symmetric contribution may not be numerically too large, although as $k_\perp$ is increased we will indeed expect the symmetric contribution to the CEP cross section to be negligible (falling as $\sim 1/k_\perp^4$). This is only a rough estimate which is based on kinematics alone, with the precise result requiring a specific calculation, but, when we recall that for example the skewed amplitude for $\pi\pi$  production is strongly suppressed by the $J_z=0$ selection rule  (by $\sim$ two orders of magnitude), it appears that the symmetric diagram may indeed be important for lower values of $M_X$, and this is equally true for vector meson production. On the other hand, for non-suppressed $\eta'\eta'$ CEP, and the relatively non-suppressed $\eta\eta'$ and $\eta\eta$ CEP, we can safely neglect the symmetric contribution.

We note that the skewed amplitude shown in Fig.~\ref{fors} (a) is almost purely imaginary while the non-perturbative amplitude in Fig.~\ref{npip} and the symmetric amplitude in Fig.~\ref{fors} (b) are almost purely real. In the case of the non-perturbative amplitude, there are two Pomeron exchanges: each Pomeron has intercept $\alpha(0)\approx 1$ and therefore gives a dominantly imaginary contribution, leading to the dominantly real product $i^2=-1$.  In the case of the symmetric perturbative diagram we have two gluon loops ($q_1\tilde q_1$ and $q_2\tilde q_2$), with each loop representing the high sub-energy, $(p_1'+l_1+l_4)^2$ or $(p_2'+l_2+l_3)^2$, positive signature amplitude, which is equivalent to a Pomeron exchange and gives an imaginary contribution; the total amplitude is therefore real.\footnote{The main features of the symmetric perturbative contribution follow closely those of the non-perturbative contribution. The perturbative amplitude of Fig.~\ref{fors} (b) may therefore be considered as the large $k_\perp$ tail of the non-perturbative amplitude.} On the other hand, the skewed diagram in Fig.~\ref{fors} (a) contains only one gluon loop (with loop momentum $Q_\perp$), which therefore gives an imaginary contribution.

We stress that the estimate of (\ref{comp}) is only approximate: there may therefore be important physics that this misses. Firstly, we have ignored the effect that including the exact form of the subprocess production amplitudes (see for instance (\ref{spec})) may have on the relative size of the skewed and symmetric contributions, although it can readily be seen that in terms of, for example, colour structure, the symmetric amplitude is not further suppressed. Secondly, and more importantly, is the issue of how to include the generalised PDFs in (\ref{bsym}). In the case of the skewed CEP amplitude shown in Fig.~\ref{fors} (a), the skewed PDFs can be related to the standard integrated (or unintegrated) PDFs (see Section~\ref{CEPform}) by using the fact that the light cone momentum fraction carried by the left screening gluon, $x'$, is much less than that, $x$, carried by the active (right) gluon, $|x'|\ll x$. On the other hand, for the symmetric case of Fig.~\ref{fors} (b) both the exchanged gluons have comparable momenta, $q_i\sim\tilde q_i$. Moreover, in contrast to the diagram for the DIS cross section, where the momentum fractions carried by $q_i$ and $\tilde q_i$ gluons have different signs (with the directions indicated as in Fig.~\ref{fors} (b)), here they have the same sign. In terms of the variables conventionally used to describe the generalised PDFs, $X=(x-x')/2$ and $\zeta=(x+x')/2$, this corresponds to the so-called `time-like' $|X|<\zeta$ domain, where the skewed parton distribution cannot be extracted from DIS data, which lie in the `space-like' domain. In the $|X|<\zeta$ domain the generalised PDFs describe the wavefunction of a `glueball' formed by two $t$-channel gluons. Since we do not know the wave (distribution) functions of the appropriate `glueballs' and their couplings to the proton, the best we can do is to put an upper limit on the cross section, based on the Schwarz inequality~\cite{Martin97}
\begin{equation}
f_g(x,x',Q^2,...)\leq\frac{1}{2}((f_g(x,-x,Q^2,...)+f_g(x',-x',Q^2,...))\;,
\end{equation}
where the diagonal gluon distribution $f_g(x,-x,Q^2,...)$ can be extracted from DIS data. Finally, we note that in the symmetric amplitude we must be careful to forbid perturbative emission from both $t$-channel exchanges, which can both be hard, although in general at a scale that is somewhat lower than $\mu=M_X/2$. We leave a full quantitative investigation of these issues to a forthcoming publication~\cite{HarlandLangfut}.

\section{Conclusions}

In this paper we have studied the central exclusive production of meson pairs, $M\overline{M}$, for a variety of light meson states at hadron colliders. We have concentrated on the CEP process in the perturbative regime, that is where the $M\overline{M}$ invariant mass is sufficiently high that the formalism outlined in Section \ref{CEPform} can be applied, with the $gg\to M\overline{M}$ subprocess described by the `hard exclusive' formalism outlined in~\cite{Brodsky81,Lepage80,Benayoun89}. In particular, the $gg\to M\overline{M}$ helicity amplitudes are calculated for a range of light scalar ($\pi^+\pi^-$, $K^+K^-$, $K^0\overline{K}^0$, $\eta\eta$, $\eta'\eta'$, $\eta\eta'$) and vector ($\rho\rho$, $\omega\omega$, $\phi\phi$) meson states, and the full $M\overline{M}$ CEP cross section is then calculated using the formalism outlined in Section \ref{CEPform}. The CEP of these states has also been included in the publicly available SuperCHIC Monte Carlo~\cite{SuperCHIC}, which generates the CEP of a range of low-mass states ($\chi_{c,b}$, $\eta_{c.b}$, $\gamma\gamma$).

We have observed and discussed various interesting theoretical properties of the $gg\to M\overline{M}$ helicity amplitudes. In Section~\ref{bg}, we have seen for example that the LO $\pi\pi$ production amplitude (and more generally that for any pair of flavour non-singlet scalar mesons, e.g. $K^+K^-$, $K^0\overline{K^0}$) vanishes when the fusing gluons are in a $J_z=0$ state. This vanishing also occurs for the production of vector mesons (e.g. $\rho\rho$, $\omega\omega$, $\phi\phi$), see Section~\ref{svect}. These results can be considered as generalisations of the vanishing of the known $\gamma\gamma \to M\overline{M}$ amplitudes for $J_z=0$ incoming photons, for neutral scalar and neutral and charged vector mesons, see~\cite{Lepage80}. We have also seen that the $|J_z|=2$ amplitude for the production of scalar mesons contains a radiation zero, where the LO amplitude vanishes exactly for a particular value of $\cos^2 \theta$ that depends on the colour factors of the $SU(3)_c$ gauge group as well as the choice of non-perturbative meson wavefunction, $\phi(x)$; this destructive interference effect will tend to suppress the production cross section. 

We have also applied the tools of the MHV formalism~\cite{Mangano90,Parke86,Berends87} to the $gg\to M\overline{M}$ calculation. In particular, the $J_z=0$ vanishing of the scalar meson amplitude, which follows from a non-trivial calculation involving many independent Feynman diagrams, has been shown to follow simply from the known Parke-Taylor amplitudes, while the $|J_z|=2$ results have been confirmed using a generalisation of MHV techniques to the calculation of non-MHV amplitudes, see~\cite{Cachazo04,Georgiou04}. While in this paper we have only explicitly calculated the amplitudes for scalar flavour non-singlet meson production within this formalism, it could readily be applied to the wider range of meson states that we have considered.

Recalling the `$J_z=0$ selection rule'~\cite{Kaidalov03,Khoze:2000mw,Khoze00a}, which strongly suppresses, by roughly two order of magnitude, the CEP cross section for non-$J_z=0$ states, the $J_z=0$ vanishing of the $gg\to M\overline{M}$ production amplitude for flavour non-singlet scalar mesons will lead to a strong suppression in the predicted CEP rate. Moreover, more generally within the hard exclusive formalism, the production cross section of two mesons at high $k_\perp$ will be strongly suppressed by the numerically small factor $(f_M/k_\perp)^4$, where $f_M\sim 100$ MeV is the meson decay constant. These two combined effects have important phenomenological consequences, for example in the case of the potential $\pi^0\pi^0$ background to the candidate $\gamma\gamma$ CEP events observed by CDF~\cite{Albrow:2010zz,cdf:2007na,Albrow}: in particular, the possible $\pi^0\pi^0$ contribution is predicted to be very small.

A further consequence of our results is that the non-resonant background to $\chi_{c0}\to \pi\pi$ CEP within this perturbative framework is predicted to be small. However, in the region $M_{\pi\pi}\sim M_\chi$ it is not completely clear that the perturbative CEP mechanism should dominate. We therefore considered in Section~\ref{nsect} the possible `non-perturbative' double Pomeron exchange contribution, calculated within the framework of Regge theory, which may give a comparable contribution at these $M_{\pi\pi}$ values, although there are some important uncertainties in the theoretical predictions at these experimentally unexplored regions of phase space. In principle, effects coming from the interference between the resonant $\chi_c$ and continuum contributions should also be considered. The perturbative tail to this non-perturbative mechanism is also considered in Section~\ref{sfors}. Although in general we find that it represents a power correction to the usual perturbative CEP process, which will be negligible in the limit of large meson $k_\perp$, in the case, for example, of the strongly suppressed $\pi\pi$ CEP cross section, we find this may also be relevant at lower values of $\pi\pi$ invariant mass. We shall consider these issues in more detail in a forthcoming publication~\cite{HarlandLangfut}.

Finally, we have also calculated the $gg\to M\overline{M}$ amplitudes for the production of flavour singlet meson states (e.g. $\eta'\eta'$), for which a new class of `ladder' diagrams contributes, see Section~\ref{eta}, with the result that the $J_z=0$ amplitudes do not vanish. From this fact we predict that the CEP cross section for the production of scalar flavour singlet states should be strongly enhanced. This also applies to the case of $\eta\eta$ and $\eta\eta'$ CEP, with the precise level of enhancement depending on the level of the $\eta-\eta'$ mixing. A further interesting possibility we have noted is that a valence $gg$ component of the flavour singlet states may further increase the cross section. In this case, there is a large level of uncertainty in the precise size of such a component, and so the CEP mechanism could in principle shed light on this issue.

\section*{Acknowledgements}

The authors thank Mike Albrow, Marius Bj\o rnstad, Stan Brodsky, Erik Brucken, Victor Chernyak, Wlodek Guryn,
Jeff Forshaw, Ronan McNulty, Dermot Moran, Jim Pinfold  and Risto Orava for useful discussions.
LHL, MGR and WJS thank the IPPP at the University of Durham for hospitality.
The work by MGR was supported  by the
grant RFBR 11-02-00120-a and by the Federal Program of the Russian
State RSGSS-65751.2010.2.
This work is also supported in part by the network PITN-GA-2010-264564
(LHCPhenoNet.). LHL acknowledges financial support from the University of Cambridge
Domestic Research Studentship scheme. WJS acknowledges financial support in the form of an IPPP Associateship.

\appendix
\renewcommand{\theequation}{A.\arabic{equation}}

\section{MHV formulae}\label{mhvf}

The spinor products are defined in terms of the solutions of the Dirac equation by
\begin{align}\label{spina}
\langle k_i\,k_j\rangle &\equiv \langle k_i^-|k_j^+\rangle = \overline{u}_-(k_i)u_+(k_j)=\overline{v}_+(k_i)v_-(k_j)\;,\\ \label{spinb}
[k_i\,k_j] &\equiv \langle k_i^+|k_j^-\rangle = \overline{u}_+(k_i)u_-(k_j)=\overline{v}_-(k_i)v_+(k_j)\;,
\end{align}
with
\begin{equation}
\langle k_i\,k_j\rangle [k_j\,k_i]=|\langle k_i\,k_j\rangle |^2=2(k_ik_j)\;.
\end{equation}
For more details and properties of the spinor products, see for example~\cite{Mangano90,Xu86}. For completeness we also reproduce the Schouten identity
\begin{equation}\label{sc}
\langle k_1\,k_2\rangle \langle k_3\,k_4\rangle =\langle k_1\,k_4\rangle \langle k_3\,k_2\rangle +\langle k_1\,k_3\rangle \langle k_2\,k_4\rangle\;, 
\end{equation}

\renewcommand{\theequation}{B.\arabic{equation}}

\section{Power-suppression of symmetric CEP mechanism: derivation.}\label{dps}

\subsection{Skewed mechanism}

The outgoing quark momenta are given by (see Fig.~\ref{fors} (a))
\begin{equation}
l_1=xk_3 \quad l_2=(1-y)k_4 \quad
l_3=yk_4 \quad l_4=(1-x)k_3\; ,
\end{equation}
where $k_{3,4}$ are the meson momenta. Considering the limit of forward outgoing protons, the incoming gluon momenta are given by
\begin{equation}
q_1=x_1p_1+Q_\perp \quad q_2=x_2p_2-Q_\perp \;.
\end{equation}
We have
\begin{equation}
k'^2=(x_1 p_1 + Q_\perp -xk_3)^2 \sim k_\perp^2\; ,
\end{equation} 
where the mesons have transverse momentum $k_\perp$. Similarly we have $k^2 \sim k_\perp^2$. The two quark propagators will therefore together give a factor of $\sim 1/k_\perp^2$ to the amplitude. Also
\begin{equation}
k_g^2=y(1-x)M_X^2\sim M_X^2\; ,
\end{equation}
where $M_X$ is the meson pair invariant mass. Finally in the amplitude we have to include the polarization vector of the incoming gluons,
\begin{equation}
\epsilon_{1\mu}\epsilon_{2\nu} \to \frac{2}{s} p_{1\mu}p_{2\nu}\to \frac{2}{M_X^2}Q_{\perp\mu} Q_{\perp\nu} \sim \frac{Q_\perp^2}{M_X^2}\;,
\end{equation}
where we have made use of the gauge invariance of the $gg\to X$ vertex $q_1^\mu V_{\mu\nu}=q_2^\mu V_{\mu\nu}=0$. Putting this together, the skewed CEP amplitude is given by
\begin{equation}
A_{\rm skew.} \sim \int\frac{{\rm d}Q_\perp^2}{Q_\perp^6}\frac{Q_\perp^2}{M_X^2}\frac{1}{k_\perp^2}\frac{1}{M_X^2}\sim \frac{1}{M_X^4}\frac{1}{k_\perp^2}\int \frac{{\rm d}Q_\perp^2}{Q_\perp^4}\;.
\end{equation}

\subsection{Symmetric mechanism}
For forward outgoing protons, the fusing gluon momenta are given by (see Fig.~\ref{fors} (b))
\begin{align}\nonumber
q_1&=x_1p_1+q_{1\perp} \quad \tilde{q}_1=\tilde{x}_1p_1-q_{1\perp}\;,\\
q_2&=x_2p_2+q_{2\perp} \quad \tilde{q}_2=\tilde{x}_2p_2-q_{2\perp}\;,
\end{align}
where $q_{1,2}$ are defined as in Fig.~\ref{fors} (b). As before we have $k'^2\sim k^2\sim k_\perp^2$ and we pick up a factor of $1/k_\perp^2$ from the quark propagators. $\vec{q}_{1\perp}$ is unconstrained and has to be integrated over, while $\vec{q}_{2\perp}$ is given by
\begin{equation}\label{q2p}
\vec{q}_{2\perp}=[x-(1-y)]\vec{k}_{\perp}-\vec{q}_{\perp}\;,
\end{equation}
apart from near the constrained region of integration in the meson wavefunctions when $x=(1-y)[1+\mathcal{O}(q_{1\perp}/k_\perp)]$. We therefore have $q_{2\perp}^2\sim k_\perp^2$, i.e. for general quark 4-momenta there has to be a large momentum transfer through both gluons $q_2$ and $\tilde{q}_2$ (or of course $q_1$ and $\tilde{q}_1$). From the gluon polarization vectors we pick up factors of $q_{1,2\perp}^2/M_X^2$, and therefore the total amplitude is given by
\begin{equation}
A_{\rm sym.}\sim \int {\rm d}q_{1_\perp}^2 \frac{1}{q_{1\perp}^4}\frac{1}{q_{2\perp}^4}\frac{q_{1\perp}^2q_{2\perp}^2}{M_X^4}\frac{1}{k_\perp^2}\sim \frac{1}{M_X^4}\frac{1}{k_\perp^4}\int\frac{{\rm d}q_{1_\perp}^2}{q_{1\perp}^2} \;.
\end{equation}
We therefore have
\begin{equation}
\frac{A_{\rm sym}}{A_{\rm skew}} \sim \frac{\langle Q_\perp^2 \rangle}{k_\perp^2}\;,
\end{equation}
where 
\begin{equation}\label{qav}
\langle Q_\perp^2 \rangle = \frac{\int {\rm d}Q_\perp^2/Q_\perp^2}{\int {\rm d}Q_\perp^2/Q_\perp^4}\;,
\end{equation}
with the usual Sudakov factor weights etc. implicit in the integrand. In general we have $\langle Q_\perp^2 \rangle=3-4\,{\rm GeV}^2$, depending on the central object mass $M_X$ and cms energy $\sqrt{s}$.

Finally, in the case of flavour-singlet meson production, for the additional diagram where the $q\overline{q}$ pairs forming the mesons are connected by a quark line, instead of (\ref{q2p}) we have
\begin{equation}
\vec{q}_{2\perp}=\vec{k}_{\perp}-\vec{q}_{\perp}\;,
\end{equation}
that is we have $q_{2\perp}^2\sim k_\perp^2$ over the full region of meson $x,y$ integration, leading to a slightly larger suppression.

D. Moran [on behalf of the LHCb collaboration], `Exclusive Dimuon Production at LHCb',
talk at DIS 2011, Newport News, USA, 11-15 April 2011.

\bibitem{rainer}
  R.~Schicker  [ALICE Collaboration],
  arXiv:1102.0742 [hep-ex].

\bibitem{Azimov74}
  Y.~I.~Azimov, V.~A.~Khoze, E.~M.~Levin {\it et al.},
  Sov.\ J.\ Nucl.\ Phys.\  {\bf 21 } (1975)  215.

\bibitem{Desai78}
  B.~R.~Desai, B.~C.~Shen and M.~Jacob,
  Nucl.\ Phys.\  B {\bf 142}, 258 (1978).

\bibitem{Lebiedowicz09}
  P.~Lebiedowicz and A.~Szczurek,
  Phys.\ Rev.\ {\bf D81}, 036003 (2010)
  [arXiv:0912.0190 [hep-ph]].

\bibitem{HarlandLang:2010ys}
  L.~A.~Harland-Lang, V.~A.~Khoze, M.~G.~Ryskin and W.~J.~Stirling,
  Eur.\ Phys.\ J.\  C {\bf 71}, 1545 (2011)
  [arXiv:1011.0680 [hep-ph]];
  arXiv:1011.1420 [hep-ph].
  
\bibitem{Brodsky81}
  S.~J.~Brodsky, G.~P.~Lepage,
  Phys.\ Rev.\  {\bf D24 } (1981)  1808.

\bibitem{Benayoun89}
  M.~Benayoun, V.~L.~Chernyak,
  Nucl.\ Phys.\  {\bf B329 } (1990)  285.

\bibitem{Lebiedowicz11}
  P.~Lebiedowicz, R.~Pasechnik, A.~Szczurek,
  [arXiv:1103.5642 [hep-ph]].

\bibitem{Staszewski:2011bg}
  R.~Staszewski, P.~Lebiedowicz, M.~Trzebinski, J.~Chwastowski, A.~Szczurek,
  
  [arXiv:1104.3568 [hep-ex]].



\bibitem{Khoze00}
  V.~A.~Khoze, A.~D.~Martin and M.~G.~Ryskin,
  Eur.\ Phys.\ J.\  C {\bf 14}, 525 (2000)
  [arXiv:hep-ph/0002072].

\bibitem{KKMRext} A.~Kaidalov, V.A.~Khoze, A.D.~Martin and M.~Ryskin, 
                  {\em Eur. Phys. J.} {\bf C 33} (2004) 261,
                  hep-ph/0311023.

 \bibitem{KMRtag} V.~A.~Khoze, A.~D.~Martin and M.~G.~Ryskin,
  Eur.\ Phys.\ J.\  C {\bf 24}, 581 (2002)
  [arXiv:hep-ph/0203122].

     \bibitem{Ryskin09}
  M.~G.~Ryskin, A.~D.~Martin and V.~A.~Khoze,
  Eur.\ Phys.\ J.\  C {\bf 60} (2009) 265
  [arXiv:0812.2413 [hep-ph]].

\bibitem{Ryskin:2011qe}
  M.~G.~Ryskin, A.~D.~Martin, V.~A.~Khoze,
  Eur.\ Phys.\ J.\  {\bf C71 } (2011)  1617.
  [arXiv:1102.2844 [hep-ph]].

\bibitem{Ryskin:2009qf}
  M.~G.~Ryskin, A.~D.~Martin, V.~A.~Khoze, A.~G.~Shuvaev,
  J.\ Phys.\ G {\bf G36 } (2009)  093001.
  [arXiv:0907.1374 [hep-ph]].

\bibitem{Chernyak06}
  V.~L.~Chernyak,
  Phys.\ Lett.\  {\bf B640 } (2006)  246-251
  [hep-ph/0605072].

\bibitem{Atkinson83}
  G.~W.~Atkinson, J.~Sucher, K.~Tsokos,
  Phys.\ Lett.\  {\bf B137 } (1984)  407.

\bibitem{Lepage80}
  G.~P.~Lepage, S.~J.~Brodsky,
  Phys.\ Rev.\  {\bf D22 } (1980)  2157.

\bibitem{Brodsky91}
  S.~J.~Brodsky, H.~C.~Pauli,
  Lect.\ Notes Phys.\  {\bf 396 } (1991)  51-121.

\bibitem{Aubert09}
  B.~Aubert {\it et al.} [ The BABAR Collaboration ],
  Phys.\ Rev.\  {\bf D80 } (2009)  052002
  [arXiv:0905.4778 [hep-ex]].

\bibitem{Druzhinin09}
  V.~P.~Druzhinin,
  PoS EPS-HEP2009  (2009)  051
  [arXiv:0909.3148 [hep-ex]].

\bibitem{Bakulev01}
  A.~P.~Bakulev, S.~V.~Mikhailov, N.~G.~Stefanis,
  Phys.\ Lett.\  {\bf B508 } (2001)  279-289
  [hep-ph/0103119].

\bibitem{Kroll96}
  P.~Kroll, M.~Raulfs,
  Phys.\ Lett.\  {\bf B387 } (1996)  848-854
  [hep-ph/9605264].

\bibitem{Radyushkin09}
  A.~V.~Radyushkin,
  Phys.\ Rev.\  {\bf D80 } (2009)  094009
  [arXiv:0906.0323 [hep-ph]].

\bibitem{Chernyak81}
  V.~L.~Chernyak, A.~R.~Zhitnitsky,
  Nucl.\ Phys.\  {\bf B201 } (1982)  492.
  
\bibitem{Chernyak09}
  V.~L.~Chernyak,
  [arXiv:0912.0623 [hep-ph]].

\bibitem{:2009cka}
  S.~Uehara {\it et al.} [ BELLE Collaboration ],
  Phys.\ Rev.\  {\bf D79 } (2009)  052009
  [arXiv:0903.3697 [hep-ex]].

\bibitem{Brodsky0s82}
  S.~J.~Brodsky, R.~W.~Brown,
  Phys.\ Rev.\ Lett.\  {\bf 49 } (1982)  966.

\bibitem{Zhu80}
  D.~-p.~Zhu,
  Phys.\ Rev.\  {\bf D22 } (1980)  2266.
\bibitem{Heyssler:1997ng}
  M.~Heyssler, W.~J.~Stirling,
  Eur.\ Phys.\ J.\  {\bf C5 } (1998)  475-484
  [hep-ph/9712314].

\bibitem{Brown95}
  R.~W.~Brown,
 AIP Conf.\ Proc.\  {\bf 350 } (1995)  261.
 [hep-th/9506018].

\bibitem{Brown82}
  R.~W.~Brown, K.~L.~Kowalski, S.~J.~Brodsky,
  Phys.\ Rev.\  {\bf D28 } (1983)  624.

\bibitem{Mangano90}
  M.~L.~Mangano, S.~J.~Parke,
  Phys.\ Rept.\  {\bf 200 } (1991)  301-367
  [hep-th/0509223].

\bibitem{Parke86}
  S.~J.~Parke, T.~R.~Taylor,
  Phys.\ Rev.\ Lett.\  {\bf 56 } (1986)  2459.

\bibitem{Berends87}
  F.~A.~Berends, W.~T.~Giele,
  Nucl.\ Phys.\  {\bf B306 } (1988)  759.

\bibitem{Mangano87}
  M.~L.~Mangano, S.~J.~Parke,
  Nucl.\ Phys.\  {\bf B299 } (1988)  673.
  
\bibitem{Mangano88}
  M.~L.~Mangano,
  Nucl.\ Phys.\  {\bf B309 } (1988)  461.

\bibitem{Cachazo04}
  F.~Cachazo, P.~Svrcek, E.~Witten,
  JHEP {\bf 0409 } (2004)  006
  [hep-th/0403047].

\bibitem{Georgiou04}
  G.~Georgiou, V.~V.~Khoze,
  JHEP {\bf 0405 } (2004)  070
  [hep-th/0404072].
  
\bibitem{Dixon96}
  L.~J.~Dixon,
  [hep-ph/9601359].

\bibitem{Birthwright05}
  T.~G.~Birthwright, E.~W.~N.~Glover, V.~V.~Khoze {\it et al.},
  JHEP {\bf 0507 } (2005)  068
  [hep-ph/0505219].

\bibitem{Wu04}
  J.~-B.~Wu, C.~-J.~Zhu,
  JHEP {\bf 0409 } (2004)  063
  [hep-th/0406146].

\bibitem{Kleiss85}
  R.~Kleiss, W.~J.~Stirling,
  Nucl.\ Phys.\  {\bf B262 } (1985)  235-262.

\bibitem{Brodsky81h}
  S.~J.~Brodsky, G.~P.~Lepage,
  Phys.\ Rev.\  {\bf D24 } (1981)  2848.

\bibitem{Nakamura10}
  K.~Nakamura {\it et al.} [ Particle Data Group Collaboration ],
  J.\ Phys.\ G {\bf G37 } (2010)  075021.

\bibitem{Thomas07}
  C.~E.~Thomas,
  JHEP {\bf 0710 } (2007)  026
  [arXiv:0705.1500 [hep-ph]].

\bibitem{Feldmann98}
  T.~Feldmann, P.~Kroll, B.~Stech,
  Phys.\ Rev.\  {\bf D58 } (1998)  114006.
  [hep-ph/9802409].

\bibitem{LY} L.D. Landau, Dokl. Akad. Nauk SSSR {\bf 60} (1948) 213;\\
C.N. Yang, Phys. Rev. {\bf 77} (1950) 242.

\bibitem{Mathieu09}
  V.~Mathieu, V.~Vento,
  Phys.\ Rev.\  {\bf D81 } (2010)  034004
  [arXiv:0910.0212 [hep-ph]].

\bibitem{Kroll02}
  P.~Kroll, K.~Passek-Kumericki,
  Phys.\ Rev.\  {\bf D67 } (2003)  054017
  [hep-ph/0210045].

\bibitem{Baier81}
  V.~N.~Baier, A.~G.~Grozin,
  Nucl.\ Phys.\  {\bf B192 } (1981)  476-488.

\bibitem{Ohrndorf81}
  T.~Ohrndorf,
  Nucl.\ Phys.\  {\bf B186 } (1981)  153.

\bibitem{Nakazawa04}
  H.~Nakazawa {\it et al.} [ BELLE Collaboration ],
  Phys.\ Lett.\  {\bf B615 } (2005)  39-49
  [hep-ex/0412058].

\bibitem{Ke11}
  H.~-W.~Ke, X.~-H.~Yuan, X.~-Q.~Li,
 [arXiv:1101.3407 [hep-ph]].

\bibitem{Ambrosino06}
  F.~Ambrosino {\it et al.} [ KLOE Collaboration ],
  Phys.\ Lett.\  {\bf B648 } (2007)  267-273.
  [hep-ex/0612029].

\bibitem{Escribano07}
  R.~Escribano, J.~Nadal,
  JHEP {\bf 0705 } (2007)  006.
  [hep-ph/0703187].

\bibitem{Martin:2009iq}
  A.~D.~Martin, W.~J.~Stirling, R.~S.~Thorne, G.~Watt,
  Eur.\ Phys.\ J.\  {\bf C63 } (2009)  189-285.
  [arXiv:0901.0002 [hep-ph]].

\bibitem{Martin:1999ww}
  A.~D.~Martin, R.~G.~Roberts, W.~J.~Stirling, R.~S.~Thorne,
  Eur.\ Phys.\ J.\  {\bf C14 } (2000)  133-145.
  [hep-ph/9907231].

\bibitem{HarlandLangfut}
 L.~A.~Harland-Lang, V.~A.~Khoze, M.~G.~Ryskin and W.~J.~Stirling, future publication.

\bibitem{SuperCHIC} The SuperCHIC code and documentation are available at {\tt http://projects.hepforge.org/superchic/}

\bibitem{Pumplin76}
  J.~Pumplin and F.~Henyey,
  Nucl.\ Phys.\  B {\bf 117}, 377 (1976).

\bibitem{regge} P.D.B. Collins, {\it Regge theory and high energy
physics}, (Cambridge Univ. Press, 1977); \\
A.C. Irving and R.P. Worden, Phys. Rept. {\bf 34}, 117 (1977).

\bibitem{DL92}
  A.~Donnachie, P.~V.~Landshoff,
  Phys.\ Lett.\  {\bf B296 } (1992)  227-232
  [hep-ph/9209205].

\bibitem{Breakstone90}
  A.~Breakstone {\it et al.} [ Ames-Bologna-CERN-Dortmund-Heidelberg-Warsaw Collaboration ],
  Z.\ Phys.\  {\bf C48 } (1990)  569-576.

\bibitem{malbrow}
 M.~Albrow, private communication.

\bibitem{Martin97}
  A.~D.~Martin, M.~G.~Ryskin,
  Phys.\ Rev.\  {\bf D57 } (1998)  6692-6700
  [hep-ph/9711371].

\bibitem{Xu86}
  Z.~Xu, D.~-H.~Zhang, L.~Chang,
  Nucl.\ Phys.\  {\bf B291 } (1987)  392.

\end{document}